\documentclass[a4paper,11pt]{article}
\pdfoutput=1 

\usepackage{jcappub} 

\usepackage[T1]{fontenc} 

\title{\boldmath Cosmological fluctuations in Delta Gravity}

\author{J. Alfaro}
\author{C. Rubio}
\author{and M. San Mart\'in}
\affiliation{Facultad de F\'isica, Pontificia Universidad Cat\'olica de Chile,\\Av. Vicu\~na Mackenna 4860, Santiago, Chile}

\emailAdd{jalfaro@fis.puc.cl}
\emailAdd{carubiof@uc.cl}
\emailAdd{mlsanmartin@uc.cl}

\abstract{About 70\% of the Universe is Dark Energy, but the physics community still does not know what it is. Delta Gravity (DG) is an alternative theory of gravitation that could solve this cosmological problem. Previously, we studied the Universe's accelerated expansion, where DG was able to explain the SNe-Ia data successfully. In this work, we explore the cosmological fluctuations that give rise to the CMB through a hydrodynamic approximation. We calculate the gauge transformations for the metric and the perfect fluid to present the equations of the evolution of cosmological fluctuations, providing the necessary equations to solve in a semi-analytical way the scalar TT Power Spectrum. These equations will be useful for comparing the DG theory with astronomical observations and thus being able to constraint the DG cosmology, testing, in the future, the compatibility with the CMB Planck data, which are currently in controversy with SNe-Ia.}

\begin{document}
\maketitle
\flushbottom

\section{Introduction}\label{sec: Introduction}

Recently there has been a spark of interest in Cosmology because the observational constraints are more precise and can constraint the physics that describes the Universe.
Despite the increasing evidence of cosmology phenomena such as the acceleration of the Universe described by the Dark Energy (DE) and the presence of a non-visible composition described by a Dark Matter (DM) composition, the Physics community has not been able to explain what they are \cite{BIB_PLANCK_Aghanim:2018eyx,Riess2016,10.1093/mnras/stx2630}.
The standard cosmological model, $\Lambda$CDM, describes the Universe composition where the DE density is 69\% of the Universe, the DM energy density is 26\%, and the rest 5\% is ordinary matter and light \cite{BIB_PLANCK_Aghanim:2018eyx}. This standard model has been able to describe the Universe using these two dark components and even can explain the SNe-Ia and CMB observations, including the formation of the large scale structure through cosmological simulations\cite{illustris,millennium}.
However, the $\Lambda$CDM model is showing inconsistencies between the early and late Universe description \cite{Addison_2018}. These problems appear in different cosmological parameters such as the Hubble constant\cite{Riess2018,Riess2019}, the curvature \cite{DiValentino2020}  or the $S_8$ tension \cite{PhysRevD.91.103508}. 
Measuring the cosmic microwave background (CMB) radiation, the Planck team found a local expansion rate of $H_0=67.37\pm 0.54$ Km/s/Mpc, which is consistent with a flat $\Lambda$ CDM model\cite{BIB_PLANCK_Aghanim:2018eyx} (where the Hubble constant must be derived taking into account other observations like BAOs). On the other hand, the SH0ES collaboration found a larger value $H_0=73.52\pm 1.62$ Km/s/Mpc through model-independent measurements of the local Universe\cite{Riess2018}, at $ \gtrsim 3.5\sigma$ discrepancy with Planck value. This tension between early and late Universe exists even without Planck CMB data or the SH0ES distance ladder\cite{Addison_2018}. Another direct measurement of $H_0=72.5^{+2.1}_{-2.3}$\;Km/s/Mpc\cite{10.1093/mnras/stz200} form the H0LiCOW collaboration based on lensing time delays is in moderate tension with Planck, while a constraint from Big Bang nucleosynthesis (BBN) combined with baryon acoustic oscillation (BAO) data of $H_0=66.98\pm 1.18$\;Km/s/Mpc\cite{Addison_2018} is inconsistent with SH0ES.

Other studies have tried to explain this discrepancy, suggesting that due to cosmic variance, the Hubble constant determined from nearby SNe-Ia may differ from that measured from the CMB by $\pm 0.8$ percent at $1 \sigma$ statistical significance. Still, this difference does not explain the discrepancy between SNe-Ia and CMB\cite{Cosmic_variance}. Nevertheless, in an extreme case, observers located in the centers of the immense voids could measure a Hubble constant from SNe-Ia biased high by 5 percent. 

From the first publication of the $H_0$ tension \cite{Riess2016} there have been many questions about the origin of this discrepancy. It has been suggested that could be errors in the calibration of Cepheids that contribute to systematic errors. This possible error has been discarded in an extensive discussion made by Riess et. al.\cite{Riess_2020}.

Other publications have tried to solve the acceleration evidence, including anisotropies at local scales. Using SNe-Ia data \cite{Wang2014} they found evidence of anisotropies associated with the direction and the amplitude of the bulk flow. Nevertheless, the effect of dipolar distribution dark energy cannot be excluded at high redshift. Also, there is another publication \cite{Evidence_anisotropy_acceleration}  where the anisotropies in cosmic acceleration are related to the Dark Energy, in their words, {\it the cosmic
acceleration deduced from supernovae may be an artifact of our being non-Copernican observers, rather than evidence for a dominant
component of "dark energy" in the Universe.}. Other studies \cite{Sun2019} conclude that even in the case of anisotropy, the Dark Energy could not be completely ruled out.  This kind of proposal could provide solutions to explain variations on the local scale, for example, different measurements on the local Hubble constant. But this kind of hypothesis could defy all the analyses made by Planck using $\Lambda$CDM model because the Dark Energy component is essential for the evolution of photons of the CMB from the last Scattering Surface until now, even more, the sum over $\Omega$ for every component in the Universe would drastically change. Many other suggestions about discrepancies have appeared, not only related to SNe-Ia measurements, but also within the Planck data itself. The anisotropies in these measurements have been debated and could be ruled out because the uncertainty tends to be very high, and the results can be very inconsistent. Even hypothesis about the possibility of a Universe with less Dark Energy \cite{Kang_2020} has appeared.

Another source of errors in the local measurement could be an inhomogeneity in the local density. \cite{Kenworthy_2019} In this scenario the presence of local structure does
not appear to impede the possibility of measuring the Hubble constant to 1\% precision, and there is no evidence of a change in the Hubble constant corresponding to an inhomogeneity.

Today,  there are different methods to obtain the Hubble constant, even with SNe-II, \cite{H0_SN_type_2}. In this research, they used SNe-II as standard candles to obtain an independent measurement of the Hubble constant. The value obtained was $H_0 = 75.8^{+5.2}_{-4.9}$ km /s/Mpc.  The local $H_0$ is higher than the value from the early Universe with a confidence level of 95\%. They concluded that {\it there is no evidence that SNe-Ia are the source of the $H_0$ tension.} 
Even, from SNe-Ia, other publication concluded, from analyzing SNe-Ia as standard candles in the near-infrared, that $H_0$ = 72.8 $\pm$ 1.6 (statistical) $\pm$2.7 (systematic) km/s/Mpc. Indeed, they concluded that the tension in the competing $H_0$ distance ladders is likely not a result of supernova systematics.

Other proposals have tried to reconcile Planck and SNe-Ia data, including modifications in the physics of the DE. In other words, introducing a \cite{InteractingDE}  equation of state of interacting dark energy component, where $w$ is allowed to vary freely, could solve the $H_0$ tension. Also, decaying dark matter model has been proposal in order to alleviate the $H_0$ and $\sigma_8$ anomalies\cite{Pandey:2019plg}; in their work they reduce the tension for both measurements when only consider Planck CMB data and the local SH0ES prior on $H_0$, however when BAOs and JLA supernova dataset are included their model is weakened.\\

Other controversies are related to inconsistencies with curvature (and other parameters needed to describe the CMB) \cite{DiValentino2020}, or are related to the tension between measurements of the amplitude of the power spectrum of density perturbations (inferred using CMB) and directly measured by large-scale structure (LSS) on smaller scales \cite{PhysRevD.91.103508}. Extension of $\Lambda$CDM models have been considered\cite{Guo_2019} trying to solve the tension of $H_0$. However, they concluded that none of these extended models can convincingly resolve the $H_0$ tension.
Through the time, the tension between Planck and SNe-Ia persist \cite{BIB_PLANCK_Aghanim:2018eyx,Riess_2020}, where the $H_0$ is the most significant tension. Furthermore, the Universe is composed principally by DE, but we still do not know what it is.

In the last decades, there have been various proposals to explain the observed acceleration of the Universe. They involve the inclusion of some additional fields in approaches like Quintessence, Chameleon, Vector Dark Energy or Massive Gravity; The addition of higher-order terms in the Einstein-Hilbert action, like $f(R)$ theories and Gauss-Bonnet terms, and the introduction of extra dimensions for a modification of gravity on large scales (\cite{DE3}). Other interesting possibilities, are the search for non-trivial ultraviolet fixed points in gravity (asymptotic safety, \cite{GR_Weinberg}) and the notion of induced gravity (\cite{induced_gravity1,induced_gravity2,induced_gravity3,induced_gravity4}). The first possibility uses exact renormalization-group techniques (\cite{Ren_group_1}-\cite{Ren_group_4}) together with lattice and numerical techniques such as Lorentzian triangulation analysis (\cite{Lorentz_triang}). Induced gravity proposes that gravitation is a residual force produced by other interactions.

Delta Gravity (DG) \cite{ALFARO2012101} is an extension of General Relativity (GR), where new fields are added to the Lagrangian by a new symmetry (for more details see \cite{ALFARO2012101,Alfaro:2012yu,Alfaro:2019qja}). This theory predicts an accelerating Universe without a cosmological constant $\Lambda$, and a Hubble parameter $H_0=74.47\pm 1.63$ Km/s/Mpc\cite{Universe5020051} when fitting SN-Ia Data, which is in agreement with SH0ES.\\
Although DG gives good results for local measurements, we need to study its cosmological predictions. In particular, the information provided by the anisotropies of matter and energy fluctuations in the Cosmic Microwave Background (CMB) could allow us to understand the physical meaning of these new fields which are included.\\
The temperature correlations give us information about the constituents of the Universe, such as baryonic and dark matter. Therefore we have to study the evolution of the CMB fluctuations from the last scattering (denoted by $t_{ls}$) to the present. Usually, these computations are carried out by codes such as CMBFast\cite{1996ApJ...469..437S,1998ApJ...494..491Z} or CAMB\footnote{http://camb.info/}\cite{Lewis_2000}, where Boltzmann equations for the fluids and its interactions provide us well-known results that are in agreement with Planck measurements\cite{BIB_PLANCK_Aghanim:2018eyx}.\\
Nevertheless, one can get a good approximation of this complex problem\cite{Mukhanov2004,weinberg2008cosmology}. In this work, we use an analytical method that consists of two steps instead of study the evolution of the scalar perturbations using Boltzmann equations. First, we use a hydrodynamic approximation, which assumes photons and baryonic plasma as a fluid in thermal equilibrium at recombination time, where there is a high rate of collisions between free electrons and photons. Second, we study the propagation of photons \cite{ALFARO2012101},   by radial geodesics from the moment when the Universe switch from opaque to transparent at time $t_{ls}$ until now.\\
In this research, we present the first steps of this essential procedure, developing the theory of scalar perturbations at first order. We discuss the gauge transformations in an extended Friedmann-Lema\^itre-Robertson-Walker (FRLW) Universe. Then we show how to get an expression for temperature fluctuations, and we demonstrate that they are gauge invariant, which is a  crucial test from a theoretical point of view. With this result, we derive a formula for the scalar contribution to temperature multipole coefficients. This formula will be useful to test the theory, and could give a sign of the physical consequence of the ``delta matter'', introduced in this theory.\\
The CMB provides cosmological constraints that are crucial to test a model. Many cosmological parameters can be obtained directly from the CMB Power Spectrum, such as $h^2\Omega_b,h^2\Omega_c,100\theta,\tau,A_s$ and $n_s$ \cite{BIB_PLANCK_Aghanim:2018eyx}, but others can be derived from constraining CMB observation with SNe-Ia or BAOs. With the study of the CMB anisotropies, we can study two aspects: the compatibility between CMB Power Spectrum and DG fluctuations and the compatibility between CMB and SNe-Ia in the DG theory.\\

The paper is organized as follows: In Section 2, we introduce the definition of DG and its equations of motion, after that we review some implications of the first law or Thermodynamics, which will allow us to interpret the physical quantities of DG. Before finishing this section, we state the ansatz that the moment of equality between matter and radiation was equal as in DG as in GR, and we discuss its implications. In section 3 we study the gauge transformation for small perturbations of both geometrical and matter fields. We choose a gauge and present the gauge-invariant equations of motion for small perturbations. In Section 4, we study the evolution of cosmological perturbations where we solve partially the equations when the Universe is dominated by radiation and when it is dominated by matter. In Section 5, we derive the formula for temperature fluctuation; here, we find that this fluctuation can be expressed in three independent and gauge invariants terms. In Section 5, we obtain a formula for temperature multipole coefficients for scalar modes. This result will allow us to test the theory with Planck CMB data in future works. Finally, we give conclusions and remarks.

\section{Definition of Delta Gravity}
In this section, we will present the action as well all the symmetries of the model and derive the equations of motion.\\
These approaches are based on the application of a variation called $\tilde{\delta}$. And it has the usual properties of a variation such as:
\begin{eqnarray}\label{deltavariation}
  \tilde{\delta}(AB)&=&(\tilde{\delta}A)B+A(\tilde{\delta}B),\nonumber\\
  \tilde{\delta}\delta A&=&\delta\tilde{\delta}A,\nonumber\\
  \tilde{\delta}(\Phi_{\mu})&=&(\tilde{\delta}\Phi)_{\mu},
  \end{eqnarray}
where $\delta$ is another variation. The main point of this variation is that, when it is applied on a field (function, tensor, etc), it produces new elements that we define as $\tilde{\delta}$ fields, which we treat them as an entirely new independent object from the original, $\tilde{\Phi}=\tilde{\delta}(\Phi)$. We use the convention that a tilde tensor is equal to the $\tilde{\delta}$ transformation of the original tensor when all its indexes are covariant.\\
Now we will present the $\tilde{\delta}$ prescription for a general action. The extension of the new symmetry is given by:
\begin{equation}
  S_0=\int d^nx{\cal L}_0(\phi,\partial_i \phi)\rightarrow S=\int d^nx({\cal L}_0(\phi,\partial_i\phi)+\tilde{\delta}{\cal L}_0(\phi,\partial_i\phi)),
\end{equation}
where $S_0$ is the original action and $S$ is the extended action in Delta Gauge Theories.\\
When we apply this formalism to the Einstein-Hilbert action of GR, we get \cite{ALFARO2012101} 
\begin{eqnarray}
\label{grav action}
S = \int d^4x \sqrt{-g} \left(\frac{R}{2\kappa} + L_M - \frac{1}{2\kappa}\left(G^{\alpha \beta} - \kappa T^{\alpha \beta}\right)\tilde{g}_{\alpha \beta} +  \tilde{L}_M\right),
\end{eqnarray}

where $\kappa = \frac{8 \pi G}{c^2}$, $\tilde{g}_{\mu \nu} = \tilde{\delta}g_{\mu \nu}$, $L_M$ is the matter Lagrangian and:

\begin{eqnarray}
\label{EM Tensor}
T^{\mu \nu} = \frac{2}{\sqrt{-g}} \frac{\delta}{\delta g_{\mu \nu}}\left[\sqrt{-g} L_M\right], \\
\label{tilde L matter}
\tilde{L}_M = \tilde{\phi}_I\frac{\delta L_M}{\delta \phi_I} + (\partial_{\mu}\tilde{\phi}_I)\frac{\delta L_M}{\delta (\partial_{\mu}\phi_I)},
\end{eqnarray}

with $\tilde{\phi}=\tilde{\delta} \phi$ are the $\tilde{\delta}$ matter fields or ``delta matter'' fields. The equations of motion are given by the variation of $g_{\mu\nu}$ and $\tilde{g}_{\mu\nu}$. It is easy to see that we get the usual Einstein's equations varying the action \eqref{grav action} with respect to $\tilde{g}_{\mu \nu}$. By the other hand, variations with respect to $g_{\mu\nu}$ give the equations for $\tilde{g}_{\mu\nu}$: 
\begin{eqnarray}
\label{tilde Eq 0}
F^{(\mu \nu) (\alpha \beta) \rho
\lambda} D_{\rho} D_{\lambda} \tilde{g}_{\alpha \beta} &+& \frac{1}{2}R^{\alpha \beta}\tilde{g}_{\alpha \beta}g^{\mu \nu} + \frac{1}{2}R\tilde{g}^{\mu \nu} - R^{\mu \alpha}\tilde{g}^{\nu}_{\alpha} - R^{\nu \alpha}\tilde{g}^{\mu}_{\alpha} + \frac{1}{2}\tilde{g}^{\alpha}_{\alpha}G^{\mu \nu} \nonumber \\
&=& \frac{\kappa}{\sqrt{-g}} \frac{\delta}{\delta g_{\mu \nu}}\left[\sqrt{-g}\left(T^{\alpha \beta}\tilde{g}_{\alpha \beta} + 2\tilde{L}_M\right)\right],
\end{eqnarray}

with:

\begin{eqnarray}
\label{F}
F^{(\mu \nu) (\alpha \beta) \rho \lambda} &=& P^{((\rho
\mu) (\alpha \beta))}g^{\nu \lambda} + P^{((\rho \nu) (\alpha
\beta))}g^{\mu \lambda} - P^{((\mu \nu) (\alpha \beta))}g^{\rho
\lambda} - P^{((\rho \lambda) (\alpha \beta))}g^{\mu \nu}\,, \nonumber \\
P^{((\alpha \beta)(\mu \nu))} &=& \frac{1}{4}\left(g^{\alpha
\mu}g^{\beta \nu} + g^{\alpha \nu}g^{\beta \mu} - g^{\alpha
\beta}g^{\mu \nu}\right),
\end{eqnarray}

where $(\mu\nu)$ denotes the totally symmetric combination of $\mu$ and $\nu$. It is possible to simplify \eqref{tilde Eq 0} (see \cite{ALFARO2012101}) to get the following system of equations:

\begin{eqnarray}
\label{Einst Eq} G^{\mu \nu} &=& \kappa T^{\mu \nu}, \\
\label{tilde Eq} F^{(\mu \nu) (\alpha \beta) \rho
\lambda} D_{\rho} D_{\lambda} \tilde{g}_{\alpha \beta} + \frac{1}{2}g^{\mu \nu}R^{\alpha \beta}\tilde{g}_{\alpha \beta} - \frac{1}{2}\tilde{g}^{\mu \nu}R &=& \kappa\tilde{T}^{\mu \nu}\,,
\end{eqnarray}
where $\tilde{T}^{\mu\nu}=\tilde{\delta}T^{\mu\nu}$. Besides, the energy momentum conservation now is given by
\begin{eqnarray}
\label{Conserv T}
D_{\nu}T^{\mu \nu} &=& 0, \\
\label{Conserv tilde T}
D_{\nu}\tilde{T}^{\mu \nu} &=& \frac{1}{2}T^{\alpha \beta}D^{\mu}\tilde{g}_{\alpha \beta} - \frac{1}{2}T^{\mu \beta} D_{\beta}\tilde{g}^{\alpha}_{\alpha} + D_{\beta}(\tilde{g}^{\beta}_{\alpha}T^{\alpha \mu}).
\end{eqnarray}
Then, we are going to work with equations \eqref{Einst Eq}, \eqref{tilde Eq}, \eqref{Conserv T} and \eqref{Conserv tilde T}. However, as the perturbation theory in the standard sector is well known, (see \cite{weinberg2008cosmology}), we will focus on DG sector.


One important result of DG is that photons follow geodesic trajectories given by the effective metric ${\bf g_{\mu\nu}}=g_{\mu\nu}+\tilde{g}_{\mu\nu}$, and for a FRLW Universe this metrics take the form

\begin{equation}\label{FRLW metric}
\bar{g}_{\mu\nu}dx^{\mu}dx^{\nu}=-dt^2+a^2(t)(dx^2+dy^2+dz^2)\,,
\end{equation}

and

\begin{equation}\label{tilde FRLW metric}
\tilde{\bar {g}}_{\mu\nu}dx^{\mu}dx^{\nu}=-3F(t)dt^2+F(t)a^2(t)(dx^2+dy^2+dz^2),
\end{equation}

where $F(t)$ is a time dependent function which is determined by the solution of the unperturbed equations system, $a(t)$ is the standard scale factor, which in Section \ref{sec:evolutionofcosmopert} we will show that it is no longer the physical scale factor of the Universe. To obtain the form of $\bar{g}_{\mu\nu}$ and $\tilde{\bar{g}}_{\mu\nu}$, first we impose isotropy and homogeneity, and then we apply the harmonic gauge $g^{\mu\nu}\Gamma^{\alpha}_{\mu\nu}=0$ and its tilde version (for details see \cite{Alfaro:2012yu}).\\
One of the implications of this effective metric is that geometry is now described by a new tridimensional metric given by\cite{ALFARO2012101}
\begin{eqnarray}\label{tridim metric}
    dl^2&=&\gamma_{ij}dx^idx^j\,,\\
    \gamma_{ij}&=&\frac{g_{00}}{{\bf g}_{00}}\left({\bf g}_{ij}-\frac{{\bf g}_{i0}{\bf g}_{j0}}{{\bf g}_{00}}\right)\,\nonumber\,,
\end{eqnarray}
while the proper time is defined by $g_{\mu\nu}$. In this case, $t$ is the cosmic time.

\subsection{DG and Thermodynamics}
Now we will study some implications of thermodynamics in cosmology for DG. Eq.\eqref{tridim metric} defines the modified scale factor of this theory:
\begin{equation}
    a_{DG}(t)=a(t)\sqrt{\frac{1+F(t)}{1+3F(t)}}\,.
\end{equation}
Then the volume of a cosmological sphere is now
\begin{equation*}
    V=\frac{4}{3}\pi r^3a_{DG}^3\,.
\end{equation*}
Any physical fluid has a density given by
\begin{equation}
    \rho=\frac{U}{c^2V}\,,
\end{equation}
where $U$ is the internal energy and $V$ is the volume. From the first law of Thermodynamics we have
\begin{equation}\label{firstlaw}
    \frac{dU}{dt}=T\frac{dS}{dt}-P\frac{dV}{dt}\,.
\end{equation}
We will assume that the Universe evolved adiabatically, this means $\dot{S}=0$. Then we get the well known relation for the energy conservation
\begin{equation} \label{energyDG}
    \dot{\rho}=-3H_{DG}\left(\rho+\frac{P}{c^2}\right)\,,
\end{equation}
with $H_{DG}=\dot{a}_{DG}/a_{DG}$.  In order to known the evolution of $\rho$ we need an equation of state $P(\rho)$. In \cite{Alfaro:2019qja} they showed that $H_{DG}(t)$ replaces the first Friedmann equation, now we know that the second Friedmann equation is the thermodynamics statement that the Universe evolves adiabatically, so the physical densities must satisfies eq. (\ref{energyDG}). If we assume $P=\omega \rho$ we found
\begin{equation}\label{DG density solution}
    \rho a_{DG}^{3(1+\omega)}=\rho_0 a_{DG\;0}^{3(1+\omega)}\,,
\end{equation}
where $\rho_0$ is the density at the present. A crucial point in this theory is that GR equations (\ref{Einst Eq}) and (\ref{Conserv T}) are valid, then we also have a similar relation for the densities of GR, but with the standard scale factor $a(t)$, explicitly
\begin{equation}\label{RG density solution}
    \rho_{GR}a^{3(1+\omega)}=\rho_{GR\;0}a_0^{3(1+\omega)}\,.
\end{equation}
Then we can relate both densities by the ratio between them
\begin{equation}
    \frac{\rho}{\rho_{GR}}\left(\sqrt{\frac{1+F(t)}{1+3F(t)}}\right)^{3(1+\omega)}= constant(\omega)\,.
\end{equation}
This ratio will be vitally important when we study the perturbations of the system. Because we will study the evolution of fractional perturbations at the last-scattering time defined as
\begin{equation}
    \delta_{GR\;\alpha}=\frac{\delta \rho_{GR\;\alpha}}{\bar{\rho}_{GR\;\alpha}+\bar{p}_{GR\;\alpha}}\,,
\end{equation}
where $\alpha$ runs between $\gamma$, $\nu$, $B$ and $D$ (photons, neutrinos, baryons and dark matter, respectively). If we consider the results from \cite{Alfaro:2019qja}, at the moment of last-scattering ($T\sim 3000\;K$) we get
\begin{equation}\label{approximation}
    \sqrt{\frac{1+F(t_{ls})}{1+3F(t_{ls})}}\sim 1\,.
\end{equation}
This mean that at that moment the physical density was proportional to the densities of GR, and without lost of generality we can take
\begin{equation}
    \delta_{phys\;\alpha}(t_{ls})=\delta_{GR\;\alpha}(t_{ls})=\delta_{\alpha}(t_{ls})\,,
\end{equation}
as it will be introduce in Section \ref{sec:evolutionofcosmopert}. In facts, eq. (\ref{approximation}) is valid for a wide range of times, from the beginning of the Universe ($z\rightarrow \infty$)  until $z\sim 10$, so this approximation is valid in the study of primordial perturbations in DG when using the equations of GR.\\
On the other hand, the number density (number of photons over the volume) at equilibrium with matter at temperature $T$ is
\begin{equation}
    n_T(\nu)d\nu=\frac{8\pi \nu^2d\nu}{e^{\frac{h\nu}{k_BT}}-1}\,;
\end{equation}
After decoupling photons travel freely from the surface of last scattering to us. So the number of photons is conserved
\begin{equation}
    dN=n_{T_{ls}}(\nu_{ls})d\nu_{ls} dV_{ls}=n_T(\nu)d\nu dV\,,
\end{equation}
as frequencies are redshifted by $\nu=\nu_{ls}a_{DG}(t_{ls})/a_{DG}$, and the volume $V=V_{ls}a_{DG}^3/a_{DG}^3(t_{ls})$ we find that in order to keep the form of a black body distribution, temperature in the number density should evolves as $T=T_{ls}a_{DG}(t_{ls})/a_{DG}$.

\subsection{Equality time \texorpdfstring{$t_{EQ}$}{TEXT} }\label{subsec: Equality moment}
After concluding this section, there is an ansatz that we need to propose in order to be completely consistent when solving the cosmological perturbation theory in the next section. This is about when the radiation was equal to the non-relativistic matter. We state that the moment when radiation and matter were equal at some $t_{EQ}$ is the same as in GR as in DG. The implication of this statement is the following: Let us consider the ratio of the matter and radiation densities of GR \eqref{RG density solution} 
\begin{eqnarray}
\frac{\rho_{GR\;M}}{\rho_{GR\;R}}=\frac{Y}{C}\,,
\end{eqnarray}
we remind that $C=\Omega_R/\Omega_M$. Then the moment of equality in GR correspond to $Y_{EQ}=C$. On the other side, if we consider the same ratio but now between the physical densities using \eqref{DG density solution}, we get  
\begin{eqnarray}
\frac{\rho_{phys;M}}{\rho_{phys\;R}}=\frac{Y_{DG}}{C_{DG}}\,,
\end{eqnarray}
where $C_{DG}=\Omega_{DG\;R}/\Omega_{DG\;M}$. Then in the equality we need to impose $Y_{DG}(Y_{EQ})=C_{DG}$, explicitly
\begin{equation}
    C_{DG}=C\frac{\sqrt{\frac{1+F(C)}{1+3F(C)}}}{\sqrt{\frac{1+F(1)}{1+3F(1)}}}\,,
\end{equation}
if we take the value from \cite{Universe5020051}, $C\sim 10^{-4}$ and $L\sim 0.45$ implies $F(C)\sim 10^{-3}<<1$ and $F(1)\sim -L/3$, then
\begin{equation}
C_{DG}=C\sqrt{\frac{1-L}{1-L/3}}\,.
\end{equation} 
This means that total density of matter and radiation today depends explicitly on the geometry measured with $L$\cite{Universe5020051}.


\section{Perturbation Theory}
Now,  we perturbed the metric as following

\begin{eqnarray}
g_{\mu\nu}=\bar g_{\mu\nu}+h_{\mu\nu}\,,\\
\tilde{g}_{\mu\nu}=\tilde{\bar{g}}_{\mu\nu}+\tilde{h}_{\mu\nu}\,.
\end{eqnarray}

Then, we follow the standard method, known as Scalar-Vector-Tensor decomposition \cite{1946ZhETF..16..587L}. This process allows us to study those sectors independently. Therefore, the perturbations are

\begin{equation}\label{standard perturbation}
  h_{00}=-E\;\;\;h_{i0}=a\left[\frac{\partial H}{\partial x^i}+G_i\right]\;\;\;h_{ij}=a^2\left[A\delta_{ij}+\frac{\partial^2 B}{\partial x^i\partial x^j}+\frac{\partial C_i}{\partial x^j}+\frac{\partial C_j}{\partial x^i}+D_{ij}\right]\,,
\end{equation}

where

\begin{equation}\label{h conditions}
\frac{\partial C_i}{\partial x^i}=\frac{\partial G_i}{\partial x^i}=0\;\;\;\frac{\partial D_{ij}}{\partial x^j}=0\;\;\;D_{ii}=0\,.
\end{equation}

This decomposition must be equivalent for $\tilde{h}_{\mu\nu}$ (by group theory):

\begin{equation}\label{tilde perturbation}
\tilde{h}_{00}=-\tilde{E}\;\;\;\tilde{h}_{i0}=a\left[\frac{\partial \tilde{H}}{\partial x^i}+\tilde{G}_i\right]\;\;\;\tilde{h}_{ij}=a^2\left[\tilde{A}\delta_{ij}+\frac{\partial^2 \tilde{B}}{\partial x^i\partial x^j}+\frac{\partial \tilde{C}_i}{\partial x^j}+\frac{\partial \tilde{C}_j}{\partial x^i}+\tilde{D}_{ij}\right]\,,
\end{equation}

with

\begin{equation}\label{th conditions}
\frac{\partial \tilde{C}_i}{\partial x^i}=\frac{\partial \tilde{G}_i}{\partial x^i}=0\;\;\;\frac{\partial \tilde{D}_{ij}}{\partial x^j}=0\;\;\;\tilde{D}_{ii}=0\,.
\end{equation}

If we replace perturbations in \eqref{Einst Eq}, \eqref{tilde Eq}, \eqref{Conserv T}, and \eqref{Conserv tilde T}, we get the equations for the perturbations. However, there are degrees of freedom that we have to take into account to have physical solutions. In the next subsection, we show how to choose a gauge to delete the nonphysical solutions.


\subsection{Choosing a gauge}

Under a space-time coordinate transformation, the metric perturbations transform as

\begin{equation}\label{gauge trans}
\Delta h_{\mu\nu}(x)=- \bar{g}_{\lambda\nu}(x)\frac{\partial \epsilon^{\lambda}}{\partial x^{\mu}}-\bar{g}_{\mu\lambda}(x)\frac{\partial \epsilon^{\lambda}}{\partial x^{\nu}}-\frac{\partial \bar{g}_{\mu\nu}}{\partial x^{\lambda}}\epsilon^{\lambda}\,.
\end{equation}

In more detail,

\begin{eqnarray}
  \Delta h_{ij}&=&-\frac{\partial \epsilon_i}{\partial x^j}-\frac{\partial \epsilon_j}{\partial x^i}+2a\dot a\delta_{ij}\epsilon_0, \label{detailed1}\\
  \Delta h_{i0}&=&-\frac{\partial \epsilon_i}{\partial t}-\frac{\partial \epsilon_0}{\partial x^i}+2\frac{\dot a}{a}\epsilon_i, \label{detailed2}\\
  \Delta h_{00}&=&-2\frac{\partial \epsilon_0}{\partial t}\,. \label{detailed3}
\end{eqnarray}

For delta perturbations we get

\begin{equation}
  \Delta \tilde{h}_{\mu\nu}=-\tilde{\bar{g}}_{\mu\lambda}\frac{\partial \epsilon^{\lambda}}{\partial x^{\nu}}-\tilde{\bar{g}}_{\lambda\nu}\frac{\partial \epsilon^{\lambda}}{\partial x^{\mu}}-\frac{\partial\tilde{\bar{g}}_{\mu\nu}}{\partial x^{\lambda}}\epsilon^{\lambda}
  -\bar{g}_{\mu\lambda}\frac{\partial \tilde{\epsilon}^{\lambda}}{\partial x^{\nu}}-\bar{g}_{\lambda\nu}\frac{\partial \tilde{\epsilon}^{\lambda}}{\partial x^{\mu}}-\frac{\partial\bar{g}_{\mu\nu}}{\partial x^{\lambda}}\tilde{\epsilon}^{\lambda}\,.
\end{equation}

In more detail,

\begin{eqnarray}
  \Delta \tilde{h}_{ij}&=&-F\frac{\partial \epsilon_i}{\partial x^j}-F\frac{\partial \epsilon_j}{\partial x^i}-\frac{\partial\tilde{\epsilon}_j}{\partial x^i}-\frac{\partial \tilde{\epsilon}_i}{\partial x^j}+\left[\epsilon_0\left(2Fa\dot a+\dot F a^2\right)+2\tilde{\epsilon}_0a\dot a\right]\delta_{ij}\;, \label{tilde detailed1}\\
  \Delta \tilde{h}_{i0}&=&-F\frac{\partial \epsilon_i}{\partial t}-3F\frac{\partial \epsilon_0}{\partial x^i}-\frac{\partial \tilde{\epsilon}_i}{\partial t}-\frac{\partial \tilde{\epsilon}_0}{\partial x^i}+2F\frac{\dot a}{a}\epsilon_i+2\frac{\dot a}{a}\tilde{\epsilon}_i\,,\label{tilde detailed2}\\
  \Delta \tilde{h}_{00}&=&-3\epsilon_0\dot F-6F\frac{\partial \epsilon_0}{\partial t}-2\frac{\partial \tilde{\epsilon}_0}{dt}\,,\label{tilde detailed3}
\end{eqnarray}

where $\epsilon$ and $\tilde{\epsilon}=\tilde{\delta}\epsilon$ defines the coordinates transformation. Also we raised and lowered index using $\bar{g}_{\mu\nu}$, so $\epsilon^0=-\epsilon_0$, $\tilde{\epsilon}^0=-\tilde{\epsilon}_0$, $\epsilon^i=a^{-2}\epsilon_i$ and $\tilde{\epsilon}^j=a^{-2}\tilde{\epsilon}_j$.\\ Following the standard procedure, we decompose the spatial part of $\epsilon^{\mu}$ and $\tilde{\epsilon}^{\mu}$ into the gradient of a spatial scalar plus a divergenceless vector:
\begin{eqnarray}\label{epsilon descomp}
  \epsilon_i&=&\partial_i\epsilon^S+\epsilon_i^V\,\,,\hspace{0.3cm}\partial_i\epsilon^V=0\,,\\
  \tilde{\epsilon}_i&=&\partial_i\tilde{\epsilon}^S+\tilde{\epsilon}_i^V\,\,,\hspace{0.3cm}\partial_i\tilde{\epsilon}^V=0\,.
\end{eqnarray}

Thus, we can compare equations \eqref{standard perturbation} and \eqref{tilde perturbation} with \eqref{detailed1}-\eqref{detailed3} and \eqref{tilde detailed1}-\eqref{tilde detailed3} to obtain the gauge transformations of the metric components:

\begin{eqnarray}\label{g trans}
\Delta A&=&\frac{2\dot a}{a}\epsilon_0\,,\;\;\;\;\Delta B=-\frac{2}{a^2}\epsilon^S\,,\nonumber\\ \Delta C_i&=&-\frac{1}{a^2}\epsilon_i^V\,,\;\;\;\;\Delta D_{ij}=0\,,\;\;\;\;\Delta E=2\dot{\epsilon}_0\,,\\ \Delta H&=&\frac{1}{a}\left(-\epsilon_0-\dot{\epsilon}^S+\frac{2\dot a}{a}\epsilon^S\right)\,,\;\;\;\;\Delta G_i=\frac{1}{a}\left(-\dot{\epsilon}_i^V+\frac{2\dot a}{a}\epsilon_i^V\right)\,,\nonumber
\end{eqnarray}

and

\begin{eqnarray}\label{delta g trans}
\Delta \tilde{A}&=&\left(\frac{2\dot{a}F}{a}+\dot{F}\right)\epsilon_0+2\frac{\dot a}{a}\tilde{\epsilon}_0\,,\;\;\;\;\Delta \tilde{B}=-\frac{2}{a^2}\left(F\epsilon^S+\tilde{\epsilon}^S\right)\,,\nonumber\\ \Delta \tilde{C}_i&=&-\frac{1}{a^2}\left(F\epsilon_i^V+\tilde{\epsilon}_i^V\right)\,,\;\;\;\;\Delta \tilde{D}_{ij}=0\,,\;\;\;\;\Delta \tilde{E}=6F\dot{\epsilon}_0+3\dot F\epsilon_0+2\dot{\tilde{\epsilon}}_0\,,\nonumber\\ \Delta \tilde{H}&=&\frac{1}{a}\left(-3F\epsilon_0-\tilde{\epsilon}_0-F\dot{\epsilon}^S-\dot{\tilde{\epsilon}}^S+\frac{2F\dot a}{a}\epsilon^S+\frac{2\dot a}{a}\tilde{\epsilon}^S\right)\,,\nonumber\\\Delta \tilde{G}_i&=&\frac{1}{a}\left(-F\dot{\epsilon}_i^V-\dot{\tilde{\epsilon}}_i^V+\frac{2F\dot a}{a}\epsilon_i^V+\frac{2\dot a}{a}\tilde{\epsilon}_i^V\right)\,.
\end{eqnarray}

There are different scenarios in which we can continue with the calculations when we impose conditions on the parameters $\epsilon_{\mu}$ and $\tilde{\epsilon}_{\mu}$. However, before discussing this, we will study the gauge transformation of energy-momentum tensors $T_{\mu\nu}$ and $\tilde{T}_{\mu\nu}$.

\subsection{\texorpdfstring{$T_{\mu\nu}$}{TEXT} and \texorpdfstring{$\widetilde{T}_{\mu\nu}$}{TEXT}}

Now we will decompose the energy-momentum tensors $T_{\mu\nu}$ and $\tilde{T}_{\mu\nu}$ in the same way. For a perfect fluid, we would have (for more details see \cite{Alfaro:2012yu})

\begin{equation}\label{Tmunu}
T_{\mu\nu}=pg_{\mu\nu}+(\rho+p)u_{\mu}u_{\nu}\,,
\end{equation}

while for $\tilde{T}_{\mu\nu}$\cite{ALFARO2012101,Alfaro:2012yu}

\begin{equation}\label{tTmunu}
\tilde{T}_{\mu\nu}=\tilde{p}g_{\mu\nu}+p\tilde{g}_{\mu\nu}+(\tilde{\rho}+\tilde{p})u_{\mu}u_{\nu}+(\rho+p)\left(\frac{1}{2}(\tilde{g}_{\mu\alpha}u_{\nu}u^{\alpha}+\tilde{g}_{\nu\alpha}u_{\mu}u^{\alpha})+u_{\mu}^Tu_{\nu}+u_{\mu}u_{\nu}^T\right)\,,
\end{equation}

where

\begin{equation}\label{norma}
g^{\mu\nu}u_{\mu}u_{\nu}=-1\,,
\end{equation}

\begin{equation}\label{norma2}
g^{\mu\nu}u_{\mu}u_{\nu}^T=0\,.
  \end{equation}

The tensors $g_{\mu\nu}$ and $\tilde{g}_{\mu\nu}$ are defined in \eqref{FRLW metric} and \eqref{tilde FRLW metric} respectively, besides we consider

\begin{eqnarray}
  p&=&\bar p+\delta p,\nonumber\\
  \rho&=&\bar{\rho}+\delta{\rho},\nonumber\\
  u_{\mu}&=&\bar{u}_{\mu}+\delta u_{\mu},\nonumber\\
  \tilde{p}&=&\tilde{\bar{p}}+\delta{\tilde{p}},\nonumber\\
  \tilde{\rho}&=&\tilde{\bar{\rho}}+\delta{\tilde{\rho}},\nonumber\\
  u_{\mu}^T&=&\bar{u}_{\mu}^T+\delta{u}_{\mu}^T\,.
\end{eqnarray}

Usually, the equation of state is given by $p(\rho)$, so we could reduce this system. For now, we will work in the generic case. When we work in the frame $\bar{u}_{\mu}=(-1,0,0,0)$ we have $\bar{u}_{\mu}^T=0$, and the normalization conditions \eqref{norma} and \eqref{norma2} give

\begin{eqnarray}
  \delta{u}^0&=&\delta{u}_0=\frac{h_{00}}{2}\nonumber\\
  \delta{u}_0^T&=&\delta{u}_T^0=0\,
\end{eqnarray}

while $\delta{u}_i$ and $\delta{u}_i^T$ are independent dynamical variables (note that $\delta{u}^{\mu}\equiv \delta(g^{\mu\nu}u_{\nu})$ is not given by $\bar{g}^{\mu\nu}\delta{u}_{\nu}$. The same is true for $\delta{u}_T^\mu$). Then, the first-order perturbation for both energy-momentum tensors ( a perfect fluid) are

\begin{equation}\label{T pert}
\delta{T}_{\mu\nu}=\bar{p}h_{\mu\nu}+\delta{p}\bar{g}_{\mu\nu}+(\bar p+\bar{\rho})(\bar{u}_{\mu}\delta{u}_{\nu}+\delta{u}_{\mu}u_{\nu})+(\delta{p}+\delta{\rho})\bar{u}_{\mu}\bar{u}_{\nu}\,,
\end{equation}

Therefore,

\begin{equation}\label{T pert det}
  \delta{T}_{ij} = \bar{p}h_{ij}+a^2\delta_{ij}\delta{p},\,,\;\;\;\delta{T}_{i0}=\bar{p}h_{i0}-(\bar{p}+\bar{\rho})\delta{u}_i,\,,\;\;\;\delta{T}_{00}=-\bar{\rho}h_{00}+\delta{\rho}\,.
\end{equation}

While

\begin{eqnarray}\label{tT pert}
  \delta{\tilde{T}}_{\mu\nu}&=&\tilde{\bar{p}}h_{\mu\nu}+\delta{\tilde{p}}\bar{g}_{\mu\nu}+\bar{p}\tilde{h}_{\mu\nu}+\delta{p}\tilde{\bar{g}}_{\mu\nu}+(\tilde{\bar{\rho}}+\tilde{\bar{p}})(\bar{u}_{\mu}\delta{u}_{\nu}+\delta{u}_{\mu}\bar{u}_{\nu})\nonumber\\&+&(\delta{\tilde{\rho}}+\delta{\tilde{p}})\bar{u}_{\mu}\bar{u}_{\nu}+(\bar{\rho}+\bar{p})\left\{\frac{1}{2}\left[\tilde{\bar{g}}_{\mu\alpha}(\bar{u}_{\nu}\delta{u}^{\alpha}+\delta{u}_{\nu}\bar{u}^{\alpha})+\tilde{h}_{\mu\alpha}\bar{u}_{\nu}\bar{u}^{\alpha}\right.\right.\nonumber\\&+&\left.\left.\tilde{\bar{g}}_{\nu\alpha}(\bar{u}_{\mu}\delta{u}^{\alpha}+\delta{u}_{\mu}\bar{u}^{\alpha})+\tilde{h}_{\nu\alpha}\bar{u}_{\mu}\bar{u}^{\alpha}\right]+\bar{u}_{\mu}^T\delta{u}_{\nu}+\delta{u}_{\mu}^T\bar{u}_{\nu}+\bar{u}_{\mu}\delta{u}_{\nu}^T+\delta{u}_{\mu}\bar{u}_{\nu}^T\right\}\nonumber\\
  &+&(\delta{\rho}+\delta{p})\left\{\frac{1}{2}\left[\tilde{\bar{g}}_{\mu\alpha}\bar{u}_{\nu}\bar{u}^{\alpha}+\tilde{\bar{g}}_{\nu\alpha}\bar{u}_{\mu}\bar{u}^{\alpha}\right]+\bar{u}_{\mu}^T\bar{u}_{\nu}+\bar{u}_{\mu}\bar{u}_{\nu}^T\right\}\,,
  \end{eqnarray}

and

\begin{eqnarray}\label{tT pert det}
  \delta{\tilde{T}}_{00}&=&-\tilde{\bar{\rho}}h_{00}-\bar{\rho}\tilde{h}_{00}+3F\delta{\rho}+\delta{\tilde{\rho}},\nonumber\\
  \delta{\tilde{T}}_{i0}&=&\tilde{\bar{p}}h_{i0}+\bar{p}\tilde{h}_{i0}-(\tilde{\bar{\rho}}+\tilde{\bar{p}})\delta{u}_i+(\bar{\rho}+\bar{p})\left\{\frac{1}{2}[Fh_{i0}-\tilde{h}_{i0}-4F\delta{u}_{i}]-\delta{u}_{i}^T\right\},\nonumber\\
  \delta{\tilde{T}}_{ij}&=&\tilde{\bar{p}}h_{ij}+\delta{\tilde{p}}a^2\delta_{ij}+\bar{p}\tilde{h}_{ij}+\delta{p}Fa^2\delta_{ij}\,,
\end{eqnarray}

where we used $\delta{u}^{\alpha}=\delta(g^{\alpha\beta}u_{\beta})=\bar{g}^{\alpha\beta}\delta{u}_{\beta}+h^{\mu\nu}\bar{u}_{\beta}$.\\ 
Generally, we decompose $\delta{u}_i\,(\delta{u}_i^T)$ into the gradient of a scalar velocity potential $\delta{u}\,(\delta{\tilde{u}})$ and a divergenceless vector $\delta{u}_i^V\,(\delta{\tilde{u}}_i^V)$, and the dissipative corrections to the inertia tensor are added as follows

\begin{eqnarray}\label{T pert det2}
  \delta{T}_{ij}&=&\bar{p}h_{ij}+a^2\left[\delta_{ij}\delta{p}+\partial_i\partial_j\pi^S+\partial_i\pi_j^V+\partial_j\pi_i^V+\pi_{ij}^T\right]\,,\\
  \delta{T}_{i0}&=&\bar{p}h_{i0}-(\bar{p}+\bar{\rho})\left(\partial_i\delta{u}+\delta{u}_i^V\right)\,,\\
  \delta{T}_{00}&=&-\bar{\rho}h_{00}+\delta{\rho}\,,
\end{eqnarray}

and

\begin{eqnarray}\label{tT pert det2}
  \delta{\tilde{T}}_{ij}&=&\tilde{\bar{p}}h_{ij}+a^2\left[\delta_{ij}\delta{\tilde{p}}+\partial_i\partial_j\tilde{\pi}^S+\partial_i\tilde{\pi}_j^V+\partial_j\tilde{\pi}_i^V+\tilde{\pi}_{ij}^T\right]+\bar{p}\tilde{h}_{ij}\nonumber\\&+&Fa^2\left[\delta_{ij}\delta{p}+\partial_i\partial_j\pi^S+\partial_i\pi_j^V+\partial_j\pi_i^V+\pi_{ij}^T\right]\,,\\
  \delta{\tilde{T}}_{i0}&=&\tilde{\bar{p}}h_{i0}+\bar{p}\tilde{h}_{i0}-(\tilde{\bar{\rho}}+\tilde{\bar{p}})(\partial_i\partial u+\partial u_i^V)\nonumber\\&+&(\bar{\rho}+\bar{p})\left\{\frac{1}{2}[Fh_{i0}-\tilde{h}_{i0}-4F(\partial_i\delta{u}+\delta{u}_i^V)]-\partial_i\delta{\tilde{u}}+\delta{\tilde{u}}_i^V\right\},\\
  \delta{\tilde{T}}_{00}&=&-\tilde{\bar{\rho}}h_{00}-\bar{\rho}\tilde{h}_{00}+3F\delta{\rho}+\delta{\tilde{\rho}}\,,
\end{eqnarray}

where $\pi_i^V\,(\tilde{\pi}_i^V)$, $\pi_{ij}^T\,(\tilde{\pi}_{ij}^T)$ and $\delta{u}_i^V\,(\delta{\tilde{u}}_i^V)$ satisfy similar conditions to  \eqref{h conditions} and \eqref{th conditions}. These conditions are (expressed before as $C_i\,(\tilde{C}_i)$, $D_{ij}\,(\tilde{D}_{ij})$ $G_i\,(\tilde{G}_i)$):
\begin{equation}\label{T and tT conditions}
\partial_i\pi_i^V=\partial_i\tilde{\pi}_i^V=\partial_i\delta{u}_i^V=\partial_i\delta{\tilde{u}}_i^V=0\,\;\;\partial_i\pi_{ij}^T=\partial_i\tilde{\pi}_{ij}^T=0\,,\;\;\pi_{ii}^T=\tilde{\pi}_{ii}^T=0\,.
\end{equation}


\subsection{Gauge Transformations for the Energy-Momentum tensors}

The gauge transformation for $T_{\mu\nu}$ is given by

\begin{equation}\label{T gauge}
\Delta \delta{T}_{\mu\nu}(x)=-\bar{T}_{\lambda\nu}(x)\frac{\partial \epsilon^{\lambda}}{\partial x^{\mu}}-\bar{T}_{\mu\lambda}(x)\frac{\partial \epsilon^{\lambda}}{\partial x^{\nu}}-\frac{\partial \bar{T}_{\mu\nu}}{\partial x^{\lambda}}\epsilon^{\lambda}\,,
\end{equation}

where the components are

\begin{eqnarray}
  \Delta \delta{T}_{ij}&=&-\bar{p}\left(\frac{\partial \epsilon_i}{\partial x^j}+\frac{\partial \epsilon_j}{\partial x^i}\right)+\frac{\partial}{\partial t}(a^2\bar{p})\delta_{ij}\epsilon_0 \label{det T gauge1},\\
  \Delta \delta{T}_{i0}&=&-\bar{p}\frac{\partial \epsilon_i}{\partial t}+\bar{\rho}\frac{\partial \epsilon_0}{\partial x^i}+2\bar{p}\frac{\dot a}{a}\epsilon_i \label{det T gauge2},\\
  \Delta \delta{T}_{00}&=&2\bar{\rho}\frac{\partial \epsilon_0}{\partial t}+\dot{\bar{\rho}}\epsilon_0\,.\label{det T gauge3}
\end{eqnarray}

While the gauge transformation of $\delta{\tilde{T}}_{\mu\nu}$ is given by

\begin{equation}\label{tT gauge}
  \Delta \delta{\tilde{T}}_{\mu\nu}=-\tilde{\bar{T}}_{\mu\lambda}\frac{\partial \epsilon^{\lambda}}{\partial x^{\nu}}-\tilde{\bar{T}}_{\lambda\nu}\frac{\partial \epsilon^{\lambda}}{\partial x^{\mu}}-\frac{\partial\tilde{\bar{T}}_{\mu\nu}}{\partial x^{\lambda}}\epsilon^{\lambda}
  -\bar{T}_{\mu\lambda}\frac{\partial \tilde{\epsilon}^{\lambda}}{\partial x^{\nu}}-\bar{T}_{\lambda\nu}\frac{\partial \tilde{\epsilon}^{\lambda}}{\partial x^{\mu}}-\frac{\partial\bar{T}_{\mu\nu}}{\partial x^{\lambda}}\tilde{\epsilon}^{\lambda}\,,
\end{equation}

where the components are

{\footnotesize
\begin{eqnarray}
  \Delta \delta{\tilde{T}}_{ij}&=&-(\tilde{\bar{p}}+\bar{p}F)\frac{\partial \epsilon_i}{\partial x^j}-(\tilde{\bar{p}}+\bar{p}F)\frac{\partial \epsilon_j}{\partial x^i}-\bar{p}\frac{\partial\tilde{\epsilon}_j}{\partial x^i}-\bar{p}\frac{\partial \tilde{\epsilon}_i}{\partial x^j}+\left[\epsilon_0\frac{\partial}{\partial t}[a^2\left(\tilde{\bar{p}}+\bar{p}F\right)]+\frac{\partial}{\partial t}(a^2\bar{p})\tilde{\epsilon}_0\right]\delta_{ij} \label{det tT gauge1}\\
  \Delta \delta{\tilde{T}}_{i0}&=&-(\tilde{\bar{p}}+\bar{p}F)\frac{\partial \epsilon_i}{\partial t}+(\tilde{\bar{\rho}}+3F\bar{\rho})\frac{\partial \epsilon_0}{\partial x^i}-\bar{p}\frac{\partial \tilde{\epsilon}_i}{\partial t}+\bar{\rho}\frac{\partial \tilde{\epsilon}_0}{\partial x^i}+2(\tilde{\bar{p}}+\bar{p}F)\frac{\dot a}{a}\epsilon_i+2\bar{p}\frac{\dot a}{a}\tilde{\epsilon}_i\label{det tT gauge2}\\
  \Delta \delta{\tilde{T}}_{00}&=&\epsilon_0\frac{\partial}{\partial t}\left(\tilde{\bar{\rho}}+3F\bar{\rho}\right)+2(\tilde{\bar{\rho}}+3F\bar{\rho})\frac{\partial \epsilon_0}{\partial t}+\dot{\bar{\rho}}\tilde{\epsilon}_0+2\bar{\rho}\frac{\partial \tilde{\epsilon}_0}{dt}\,.\label{det tT gauge3}
\end{eqnarray}}

$\epsilon_i$ and $\tilde{\epsilon}_i$ were decomposed in \eqref{epsilon descomp} to write these gauge transformations in terms of the scalar, vector and tensor components. The transformations \eqref{detailed1}-\eqref{detailed3} and \eqref{tilde detailed1}-\eqref{tilde detailed3} with \eqref{det T gauge1}-\eqref{det T gauge2} and \eqref{det tT gauge1}-\eqref{det tT gauge3} give the gauge transformation for the pressure, energy density and velocity potential:

\begin{equation}\label{T results}
\Delta \delta{p}=\dot{\bar{p}}\epsilon_0\,,\;\;\;\Delta \delta{\rho}=\dot{\bar{\rho}}\epsilon_0\;,\;\;\;\Delta\delta{u}=-\epsilon_0\,.
\end{equation}

The other ingredients of the energy-momentum tensor are gauge invariants:

\begin{equation}
\Delta \pi^S=\Delta\pi_i^V=\Delta\pi_{ij}^T=\Delta\delta{u}_i^V=0\,.
\end{equation}

Nevertheless, the other transformations are
\begin{subequations}
\begin{eqnarray}
  \Delta\delta{\tilde{\rho}}&=&\frac{\partial}{\partial t}(\tilde{\bar{\rho}}+3F\bar{\rho})\epsilon_0+2(\tilde{\bar{\rho}}+3F\bar{\rho})\dot{\epsilon_0}+\dot{\bar{\rho}}\tilde{\epsilon}_0+2\bar{\rho}\dot{\tilde{\epsilon}}_0-\tilde{\bar{\rho}}\Delta{E}\nonumber\\&-&3F\bar{\rho}\Delta{\tilde{E}}-3F\Delta{\delta{\rho}}\,,\\
  \Delta\delta{\tilde{p}}&=&\frac{1}{a^2}\frac{\partial}{\partial t}[a^2\left(\tilde{\bar{p}}+\bar{p}F\right)]\epsilon_0+\frac{1}{a^2}\frac{\partial}{\partial t}(a^2\bar{\rho})\tilde{\epsilon}_0-\tilde{\bar{p}}\Delta{A}-\bar{p}F\Delta{\tilde{A}}-F\Delta{\delta{p}}\,,\\
  \Delta{\delta{\tilde{u}}}&=&\frac{1}{(\bar{\rho}+\bar{p})}\left\{(\tilde{\bar{p}}+\bar{p}F)\dot{\epsilon}^S-(\tilde{\bar{\rho}}+3F\bar{\rho})\epsilon_0+\bar{p}\dot{\tilde{\epsilon}}^S-\bar{\rho}\tilde{\epsilon}_0-2(\tilde{\bar{p}}+\bar{p}F)\frac{\dot{a}}{a}\epsilon^S\right.\nonumber\\&-&\left.2\bar{p}\frac{\dot{a}}{a}\tilde{\epsilon}^S+\tilde{\bar{p}}a\Delta{H}+\bar{p}a\Delta{\tilde{H}}-(\bar{\rho}+\bar{p})\left[\frac{1}{2}(1-F)a\Delta{\tilde{H}}+2F\Delta{\delta{u}}\right]\right\}\,,\\
  \Delta\delta{\tilde{u}}_i^V&=&\frac{1}{(\bar{\rho}+\bar{p})}\left\{(\tilde{\bar{p}}+\bar{p}F)\dot{\epsilon}_i^V+\bar{p}\dot{\tilde{\epsilon}}_{i}^V-2(\tilde{\bar{p}}+\bar{p}F)\frac{\dot a}{a}\epsilon_i^V-2\bar{p}\frac{\dot a}{a}\tilde{\epsilon}_i^V+\tilde{\bar{p}}a\Delta{G}_i\right.\nonumber\\&+&\left.\bar{p}a\Delta{\tilde{G}}_i-\frac{1}{2}(\bar{\rho}+\bar{p})(1-F)a\Delta{\tilde{G}}_i\right\}\,,\\
  \Delta{\delta{\tilde{\pi}}}^S&=&-\frac{2}{a^2}(\tilde{\bar{p}}+\bar{p}F)\epsilon^S-2\frac{\bar{p}}{a^2}\tilde{\epsilon}^S-\tilde{\bar{p}}\Delta{B}-\bar{p}F\Delta{\tilde{B}}\,,\\
  \Delta{\delta{\tilde{\pi}}}_i^V&=&-\frac{1}{a^2}(\tilde{\bar{p}}+\bar{p}F)\epsilon_i^V-\frac{\bar{p}}{a^2}\tilde{\epsilon}_i^V-\tilde{\bar{p}}\Delta{C}_i-\bar{p}F\Delta{\tilde{C}}_i\,,\\
  \Delta{\delta{\tilde{\pi}}}_{ij}&=&0\,.
\end{eqnarray}
\end{subequations}
The results given in \eqref{g trans}, \eqref{delta g trans} and \eqref{T results} are used to obtain

\begin{subequations}
\begin{eqnarray}
  \Delta{\delta{\tilde{\rho}}}&=&\dot{\tilde{\bar{\rho}}}\epsilon_0+(\dot{\bar{\rho}}-3F\bar{\rho})\tilde{\epsilon}_0\,,\\
  \Delta{\delta{\tilde{p}}}&=&\dot{\tilde{\bar{p}}}\epsilon_0+\dot{\bar{p}}\tilde{\epsilon}_0\,,\\
  \Delta{\delta{\tilde{u}}}&=&\left[(1-3F)\frac{F}{2}-\frac{(\tilde{\bar{p}}+\tilde{\bar{\rho}})}{(\bar{p}+\bar{\rho})}\right]\epsilon_0-\frac{1}{2}(1+F)\tilde{\epsilon}_0-(1-F)\frac{\dot{a}}{a}\left(F\epsilon^S+\tilde{\epsilon}^S\right)\nonumber\\&+&\frac{1}{2}(1-F)\left(F\dot{\epsilon}^S+\dot{\tilde{\epsilon}}^S\right)\,,\\
  \Delta{\delta{\tilde{u}}}_i^V&=&\frac{1}{2}(1-F)[F\dot{\epsilon}_i^V+\dot{\tilde{\epsilon}}_i^V-2\frac{\dot a}{a}F\epsilon_i^V-2\frac{\dot a}{a}\tilde{\epsilon}_i^V]\,,\\  
  \Delta{\delta{\tilde{\pi}}}_i^S&=&0\,,\\
  \Delta{\delta{\tilde{\pi}}}_i^V&=&0\,,\\
  \Delta{\delta{\tilde{\pi}}}_{ij}&=&0\,.
\end{eqnarray}
\end{subequations}

As we said before, there are different choices for $\epsilon$ and $\tilde{\epsilon}$ parameter to fix all the gauge freedoms. The most common and well-known gauges are the Newtonian gauge and Synchronous gauge. The former fix $\epsilon^S$ such that $B=0$, and choose $\epsilon_0$ such that $H=0$ ( in equation \eqref{g trans} ).  In DG, this choice is extended imposing similar conditions in \eqref{delta g trans} for $\tilde{\epsilon}^S$ and $\tilde{\epsilon}_0$, such that $\tilde{B}=\tilde{H}=0$. There is no remaining freedom to make a gauge transformation in this scenario. Nevertheless, in this work, we will use the Synchronous gauge, where we will choose $\epsilon_0$ such that $E=0$, and $\epsilon^S$ such that $H=0$, (similar conditions for $\tilde{\epsilon}_0$ and $\tilde{\epsilon}^S$). In the next section, we present the perturbed equations of motion in this frame, and we discuss the suitability of this choice for our purposes.


\subsection{Fields equations and energy momentum conservations in synchronous gauge}

Under this gauge fixing, perturbed Einstein equations Eq. (\ref{Einst Eq}) reads (at first order):

\begin{equation}
-4\pi G(\delta \rho+3\delta p+\nabla^2\pi^S)=\frac{1}{2}\left(3\ddot{A}+\nabla^2\ddot{B}\right)+\frac{\dot{a}}{2a}\left(3\dot{A}+\nabla^2\dot{B}\right).
\end{equation}

While the energy-momentum conservation gives

\begin{eqnarray}
  \delta p+\nabla^2\pi^S+\partial_0[(\bar{\rho}+\bar{p})\delta u]+\frac{3\dot{a}}{a}(\bar{\rho}+\bar{p})\delta u&=&0,\\
  \delta\dot{\rho}+\frac{3\dot{a}}{a}(\delta\rho+\delta p)+\nabla^2\left[a^{-2}(\bar{\rho}+\bar{p})\delta u+\frac{\dot{a}}{a}\pi^S\right]
  +\frac{1}{2}(\bar{\rho}+\bar{p})\partial_0\left[3A+\nabla^2B\right]&=&0.
  \end{eqnarray}
  
We define 

\begin{equation}
\Psi\equiv\frac{1}{2}\left[3{A}+\nabla^2{B}\right],
\end{equation}

then,

\begin{equation}\label{einst eq}
-4\pi Ga^2(\delta \rho+3\delta p+\nabla^2\pi^S)=\frac{\partial}{\partial t}\left(a^2\dot{\Psi}\right),
\end{equation}

\begin{equation}\label{e-m eq}
  \delta\dot{\rho}+\frac{3\dot{a}}{a}(\delta\rho+\delta p)+\nabla^2\left[a^{-2}(\bar{\rho}+\bar{p})\delta u+\frac{\dot{a}}{a}\pi^S\right] +\frac{1}{2}(\bar{\rho}+\bar{p})\dot{\Psi}=0.
\end{equation}

The unperturbed Einstein equations correspond to the Friedmann equations. In the Delta sector, computations give the non-perturbed equations:

\begin{equation}\label{tfriedman eq1}
3\dot{F}\frac{\dot{a}}{a}=\kappa(3F\bar{\rho}+\tilde{\bar{\rho}})
\end{equation},
and
\begin{equation}\label{tfriedman eq2}
  12F\frac{\ddot{a}}{a}+6F\left(\frac{\dot{a}}{a}\right)^2+3\dot{F}\frac{\dot{a}}{a}-3\ddot{F}=\kappa(\tilde{\bar{\rho}}+3\tilde{\bar{p}}+3F\bar{\rho}+3F\bar{p}).
\end{equation}

The perturbed contribution (at first order) is

\begin{eqnarray}\label{tilde pert eqs}
\left[2\dot{F}\frac{\dot{a}}{a}+\ddot{F}\right]\left[3A+\nabla^2B\right]+\left[6F\frac{\dot{a}}{a}+\frac{5}{2}\dot{F}\right]\left[3\dot{A}+\nabla^2\dot{B}\right]\nonumber-\left[2\frac{\dot{a}}{a}\right]\left[3\dot{\tilde{A}}+\nabla^2\dot{\tilde{B}}\right]\\+3F\left[3\ddot{A}+\nabla^2\ddot{B}\right]-\left[3\ddot{\tilde{A}}+\nabla^2\ddot{\tilde{B}}\right]=\kappa\left(3\delta{\tilde{p}}+\delta{\tilde{\rho}}+F\delta\rho+3F\delta p+\nabla^2\tilde{\pi}+F\nabla^2\pi\right)
\end{eqnarray}

Besides, $00$ component of delta Energy-momentum conservation in \eqref{Conserv tilde T} give

\begin{eqnarray}\label{tilde energy eq}
  \delta\dot{\tilde{\rho}}+\frac{3\dot{a}}{a}(\delta{\tilde{\rho}}+\delta{\tilde{p}})+\frac{3\dot{F}}{2}(\delta{\rho}+\delta{p})+\nabla^2\left[\frac{(\tilde{\bar{\rho}}+\tilde{\bar{p}})}{a^2}\delta{u}+\frac{(\bar{\rho}+\bar{p})F}{a^2}\delta{u}+\frac{(\bar{\rho}+\bar{p})}{a^2}\delta{\tilde{u}}\right]\nonumber\\+\frac{(\tilde{\bar{\rho}}+\tilde{\bar{p}})}{2}\partial_0[3A+\nabla^2B]+\frac{(\bar{\rho}+\bar{p})}{2}\partial_0[3\tilde{A}+\nabla^2\tilde{B}]-\frac{(\bar{\rho}+\bar{p})}{2}\partial_0(F[3A+\nabla^2B])=0\,,
\end{eqnarray}

while the $i0$ component gives

\begin{eqnarray}\label{tilde momentum eq}
 \delta{\tilde{p}}+\partial_0[(\tilde{\bar{\rho}}+\tilde{\bar{p}})\delta{u}]+\partial_0[(\bar{\rho}+\bar{p})\delta{\tilde{u}}]-\partial_0[(\bar{\rho}+\bar{p})F\delta{u}]+3(\bar{\rho}+\bar{p})\dot{F}\delta{u}\nonumber\\+\frac{3\dot{a}}{a}(\bar{\rho}+\bar{p})\delta{\tilde{u}}+\frac{3\dot{a}}{a}(\tilde{\bar{\rho}}+\tilde{\bar{p}})\delta{u}-\frac{3\dot{a}}{a}F(\bar{\rho}+\bar{p})\delta{u}=0\;.
\end{eqnarray}

Analogous to the standard sector, we define

\begin{equation}
\tilde{\Psi}\equiv\frac{1}{2}\left[3\tilde{A}+\nabla^2\tilde{B}\right]\;,
\end{equation}

then the gravitational equation becomes

\begin{eqnarray}\label{final DG grav eq}
\left[2\dot{F}\frac{\dot{a}}{a}+\ddot{F}\right]a^2\Psi+\left[6F\frac{\dot{a}}{a}+\frac{5}{2}\dot{F}\right]a^2\dot{\Psi}+3Fa^2\ddot{\Psi}-\frac{d}{dt}\left(a^2\dot{\tilde{\Psi}}\right)=\frac{\kappa}{2}\left(3\delta{\tilde{p}}+\delta{\tilde{\rho}}+F\delta\rho\right.\nonumber\\+\left.3F\delta p+\nabla^2\tilde{\pi}+F\nabla^2\pi\right).
\end{eqnarray}

Now, the delta energy conservation is given by

\begin{eqnarray}\label{final tilde energy eq}
  \delta\dot{\tilde{\rho}}+\frac{3\dot{a}}{a}(\delta{\tilde{\rho}}+\delta{\tilde{p}})+\frac{3\dot{F}}{2}(\delta{\rho}+\delta{p})+\nabla^2\left[\frac{(\tilde{\bar{\rho}}+\tilde{\bar{p}})}{a^2}\delta{u}+\frac{(\bar{\rho}+\bar{p})F}{a^2}\delta{u}+\frac{(\bar{\rho}+\bar{p})}{a^2}\delta{\tilde{u}}\right]\nonumber\\+(\tilde{\bar{\rho}}+\tilde{\bar{p}})\dot{\Psi}+(\bar{\rho}+\bar{p})\dot{\tilde{\Psi}}-(\bar{\rho}+\bar{p})\partial_0(F\Psi)=0\,.
\end{eqnarray}

The study of the non-perturbed sector was already treated in Alfaro et al. and applied to the supernovae observations. 
We will consider these results when necessary. For now, we only need the expression for the time dependent function $F(t)$, which is
\begin{equation}\label{F function}
F(Y) = -\frac{LY}{3}\sqrt{Y+C}\,,
\end{equation}

where $Y\equiv Y(t)=a(t)/a_0$ is the quotient between the scale factor at a time $t$ over the scale factor in the actuality (which for our purposes we will consider equal to one). $L$ $(\sim 0.45)$ and $C$ $(\sim 10^{-4})$ are the new parameters of DG that are already determined by supernova data\cite{Alfaro:2012yu,Universe5020051}. We have to remark that our definition of $\Psi$ is not the usual since the standard definition is with the time derivative of fields $A$ and $B$, respectively. In the delta sector appears explicitly the combinations of these fields without a time derivative, so if the reader wants to compare results with other works, he or she should take into consideration this definition to analyze the gauge. In the next section, we will discuss the evolution of the cosmological fluctuations, which will help us compute the scalar contribution to the CMB.


\section{Evolution of cosmological fluctuations}\label{sec:evolutionofcosmopert}

Until now, we have developed the perturbation theory in DG; now, we are interested in studying the evolution of the cosmological fluctuations to have a physical interpretation of the delta matter fields, which this theory naturally introduces. Even in the standard cosmology, the system of equations that describes these perturbations are complicated to allow analytic solutions, and there are comprehensive computer programs to this task, such as CMBfast\cite{1996ApJ...469..437S,1998ApJ...494..491Z}, and CAMB\cite{Lewis_2000}. However, such computer programs can not give a clear understanding of the physical phenomena involved. Nevertheless, some good approximations allow to compute the spectrum of the CMB fluctuations with a rather good agreement with these computer programs\cite{weinberg2008cosmology,Mukhanov2004}. In particular, we are going to extend Weinberg approach for this task. This method consists of two main aspects: first, the hydrodynamic limit, which assumes that near recombination time photons were in local thermal equilibrium with the baryonic plasma, then photons could be treated hydro-dynamically, like plasma and cold dark matter. Second, a sharp transition from thermal equilibrium to complete transparency at the moment $t_{ls}$ of the last scattering. \\
 Since we will reproduce this approach, we consider the Universe's standard components, which means photons, neutrinos, baryons, and cold dark matter. Then the task is to understand the role of their own delta-counterpart. We will also neglect both anisotropic inertia tensors and took the usual state equation for pressures and energy densities and perturbations. Besides, as we will treat photons and delta photons hydro-dynamically, we will use $\delta u_{\gamma}=\delta u_B$ and $\delta \tilde{u}_{\gamma}=\delta \tilde{u}_B$. Finally, as the synchronous scheme does not completely fix the gauge freedom, one can use the remaining freedom to put $\delta u_D=0$, which means that cold dark matter evolves at rest with respect to the Universe expansion. In our theory, the extended synchronous scheme also has extra freedom, which we will use to choose $\delta\tilde{u}_D=0$ as its standard part. Now we will present the equations for both sectors. However, we will provide more detail in the delta sector because Weinberg\cite{weinberg2008cosmology} already calculates the solution of Einstein's equations.\\ Einstein's equations and its energy-momentum conservation in Fourier space are

 \begin{eqnarray}
   \frac{d}{dt}\left(a^{2}\dot{\Psi}_{q}\right)&=&-4\pi Ga^{2}\left(\delta\rho_{Dq}+\delta\rho_{Bq}+2\delta\rho_{\gamma q}+2\delta\rho_{\nu q}\right)\,,\label{ein eq1}\\
   \delta\dot{\rho}_{\gamma q}+4H\delta\rho_{\gamma q}-(4q/3a)\bar{\rho}_{\gamma}\delta u_{\gamma q} & = & -(4/3)\bar{\rho}_{\gamma}\dot{\Psi}_{q}\,,\\
\delta\dot{\rho}_{Dq}+3H\delta\rho_{Dq} & = & -\bar{\rho}_{D}\dot{\Psi}_{q}\,,\\
\delta\dot{\rho}_{Bq}+3H\delta\rho_{Bq}-(q/a)\bar{\rho}_{B}\delta u_{\gamma q} & = & -\bar{\rho}_{B}\dot{\Psi}_{q}\,,\\
\delta\dot{\rho}_{\nu q}+4H\delta\rho_{\nu q}-(4q/3a)\bar{\rho}_{\nu}\delta u_{\nu q} & = & -(4/3)\bar{\rho}_{\nu}\dot{\Psi}_{q}\,,\label{ein eq2}
 \end{eqnarray}
 
 where $H\equiv \dot{a}/a$. It is useful to rewrite these equations in term of the dimensionless fractional perturbation
 
 \begin{equation}\label{dim frac pert}
\delta_{\alpha q}=\frac{\delta \rho_{\alpha q}}{\bar{\rho}_{\alpha}+\bar{p}_{\alpha}}\,,
   \end{equation}
 
 where $\alpha$ can be $\gamma$, $\nu$, $B$ and $D$ (photons, neutrinos, baryons and dark matter, respectively). $a^4\bar{\rho}_{\gamma}$, $a^4\bar{\rho}_{\nu}$, $a^3\bar{\rho}_{D}$, $a^3\bar{\rho}_{B}$ are time independent quantities, then  \eqref{ein eq1}-\eqref{ein eq2} are
 
 \begin{subequations}
 \begin{eqnarray}\label{GRequations}
\frac{d}{dt}\left(a^{2}\dot{\Psi}_{q}\right) & = & -4\pi Ga^{2}\left(\bar{\rho}_{D}\delta_{Dq}+\bar{\rho}_{B}\delta_{Bq}+\frac{8}{3}\bar{\rho}_{\gamma}\delta_{\gamma q}+\frac{8}{3}\bar{\rho}_{\nu}\delta{}_{\nu q}\right)\,,\label{final eq1}\\
\dot{\delta}_{\gamma q}-(q^2/a^{2})\delta u_{\gamma q} & = & -\dot{\Psi}_{q}\,,\label{final fot}\\
\dot{\delta}_{Dq} & = & -\Psi_{q}\,,\\
\dot{\delta}_{Bq}-(q^2/a^{2})\delta u_{\gamma q} & = & -\dot{\Psi}_{q}\,,\label{final bar}\\
\dot{\delta}_{\nu q}-(q^2/a^{2})\delta u_{\nu q} & = & -\dot{\Psi}_{q}\,,\\
\frac{d}{dt}\left(\frac{\left(1+R\right)\delta u_{\gamma q}}{a}\right) & = & -\frac{1}{3a}\delta_{\gamma q}\,,\\
\frac{d}{dt}\left(\frac{\delta u_{\nu q}}{a}\right) & = & -\frac{1}{3a}\delta_{\nu q}\,,\label{final eq2}
\end{eqnarray}
\end{subequations}

 where $R=3\bar{\rho}_B/4\bar{\rho}_{\gamma}$. By the other side, in delta sector we will use a dimensionless fractional perturbation. However, this perturbation is defined as the delta transformation of \eqref{dim frac pert}\footnote{We choose this definition because the system of equations now seems as an homogeneous system exactly equal to the GR sector (where now the variables are the tilde-fields) with external forces mediated by the GR solutions. Maybe the most intuitive solution should be $$\tilde{\delta}^{int}_{\alpha q}=\frac{\delta\tilde{\rho}_{\alpha q}}{\tilde{\bar{\rho}}_{\alpha}+\tilde{\bar{p}}_{\alpha}}\;,$$ however these definitions are related by $$\tilde{\delta}_{\alpha q}=\frac{\tilde{\bar{\rho}}_{\alpha}+\tilde{\bar{p}}_{\alpha}}{\bar{\rho}_{\alpha}+\bar{p}_{\alpha}}\left(\tilde{\delta}^{int}_{\alpha q}-\delta_{\alpha q}\right)\;.$$},
 
 \begin{equation}\label{tilde dim frac pert}
\tilde{\delta}_{\alpha q}\equiv \tilde{\delta} \delta_{\alpha q}=\frac{\delta \tilde{\rho}_{\alpha q}}{\bar{\rho}_{\alpha}+\bar{p}_{\alpha}}-\frac{\tilde{\bar{\rho}}_{\alpha}+\tilde{\bar{p}}_{\alpha}}{\bar{\rho}_{\alpha}+\bar{p}_{\alpha}}\delta_{\alpha q}\,.
 \end{equation}
 
 In \cite{Alfaro:2012yu}, they found
 
 \begin{equation} \label{density solution}
\frac{\tilde{\bar{\rho}}_R}{\bar{\rho}_R}=-2F(a)\;\;\;\;and\;\;\;\;\frac{\tilde{\bar{\rho}}_M}{\bar{\rho}_M}=-\frac{3}{2}F(a)\,.
\end{equation}

 We will assume that this quotient holds for every component. Also using the result that  $a^4\tilde{\bar{\rho}}_{\gamma}/F$, $a^4\tilde{\bar{\rho}}_{\nu}/F$, $a^3\tilde{\bar{\rho}}_{D}/F$, $a^3\tilde{\bar{\rho}}_{B}/F$ are time independent, the equations for the delta sector are
 
\begin{subequations}
 \begin{eqnarray} \label{fdelta eq1}
\left[2\dot{F}\frac{\dot{a}}{a}+\ddot{F}\right]a^2\Psi_q+\left[6F\frac{\dot{a}}{a}+\frac{5}{2}\dot{F}\right]a^2\dot{\Psi}_q+3Fa^2\ddot{\Psi}_q-\frac{d}{dt}\left(a^2\dot{\tilde{\Psi}}_q\right)\nonumber\\=\frac{\kappa}{2} a^2\left[\bar{\rho}_D\tilde{\delta}_{Dq}+\bar{\rho}_B\tilde{\delta}_{Bq}+\frac{8}{3}\bar{\rho}_{\gamma}\tilde{\delta}_{\gamma q}+\frac{8}{3}\bar{\rho}_{\nu}\tilde{\delta}_{\nu q}-\frac{F}{2}\left(\bar{\rho}_D{\delta}_{Dq}+\bar{\rho}_B{\delta}_{Bq}\right)\right.&-&\left.\frac{8}{3}F\left(\bar{\rho}_{\gamma}{\delta}_{\gamma q}+\bar{\rho}_{\nu}{\delta}_{\nu q}\right)\right]\,,\nonumber\\
\dot{\tilde{\delta}}_{\gamma q}-\frac{q^2}{a^2}\left(\delta\tilde{u}_{\gamma q}+F\delta u_{\gamma q}\right)+\dot{\tilde{\Psi}}_q-\partial_0(F\Psi_q)&=&0\;,\label{final delta fot}\\
\dot{\tilde{\delta}}_{Dq}+\dot{\tilde{\Psi}}_q-\partial_0(F\Psi_q)&=&0\;,\\
\dot{\tilde{\delta}}_{Bq}-\frac{q^2}{a^2}\left(\delta\tilde{u}_{\gamma q}+F\delta u_{\gamma q}\right)+\dot{\tilde{\Psi}}_q-\partial_0(F\Psi_q)&=&0\;,\label{final delta B}\\
\dot{\tilde{\delta}}_{\nu q}-\frac{q^2}{a^2}\left(\delta\tilde{u}_{\nu q}+F\delta u_{\nu q}\right)+\dot{\tilde{\Psi}}_q-\partial_0(F\Psi_q)&=&0\;,\\
\frac{\tilde{\delta}_{\gamma q}}{3a}+\frac{d}{dt}\left(\frac{(1+R)\delta \tilde{u}_{\gamma q}}{a}\right)+2F\frac{d}{dt}\left(\frac{(R-\tilde{R})\delta {u}_{\gamma q}}{a}\right)\nonumber\\-F\frac{d}{dt}\left(\frac{(1+R)\delta u_{\gamma q}}{a}\right)-2\dot{F}(\tilde{R}-R)\frac{\delta u_{\gamma q}}{a}&=&0\;,\\
\frac{\tilde{\delta}_{\nu q}}{3a}+\frac{d}{dt}\left(\frac{\delta\tilde{u}_{\nu q}}{a}\right)-F\frac{d}{dt}\left(\frac{\delta u_{\nu q}}{a}\right)&=&0\;, \label{fdelta eq2}
\end{eqnarray}
\end{subequations}
 with $\tilde{R}=3\tilde{\bar{\rho}}_B/4\tilde{\bar{\rho}}_{\gamma}$. Due to the definition of tilde fractional perturbation \eqref{tilde dim frac pert}, solutions for \eqref{fdelta eq1}-\eqref{fdelta eq2} can be obtained easily, putting all solutions of GR equal to zero, then the system is exactly equal to the system of equations \eqref{final eq1}-\eqref{final eq2} and the solution of tilde perturbations in the homogeneous system are exactly equal to the GR solutions, and then we only need to "turn on" the GR source and find the complete solutions just like a forced-system.\\
 We will impose initial conditions to find solutions valid up to recombination time. At sufficiently early times the Universe was dominated by radiation, and as Friedmann equations are valid in our theory (in particular the first equation), we can use a good approximation given by $a\propto \sqrt{t}$ and $8\pi G\bar{\rho}_{R}/3=1/4t^2$, while $R$ and $\tilde{R} \ll 1$. Here
 
 \begin{equation}\label{total densities}
   \bar{\rho}_M\equiv \bar{\rho}_D+\bar{\rho}_B\;,\;\;\;\;\bar{\rho}_R\equiv \bar{\rho}_{\gamma}+\bar{\rho}_{\nu}\,.
 \end{equation}

 Besides, we are interested in adiabatic solutions, in the sense that all the $\delta_{\alpha q}$ and $\tilde{\delta}_{\alpha q}$ become equal at very early times. So, we make the ansatz:
 
 \begin{eqnarray}\label{adiabaticity}
   \delta_{\gamma q}=\delta_{\nu q}=\delta_{B q}=\delta_{D q}=\delta_q\,,\;\;\;\;\delta u_{\gamma q}=\delta u_{\nu q}=\delta u_q\;,\\
   \tilde{\delta}_{\gamma q}=\tilde{\delta}_{\nu q}=\tilde{\delta}_{B q}=\tilde{\delta}_{D q}=\tilde{\delta}_q\,,\;\;\;\;\delta \tilde{u}_{\gamma q}=\delta \tilde{u}_{\nu q}=\delta \tilde{u}_q\,.
   \end{eqnarray}
   
Finally, we drop the term $q^2/R^2$ because we are considering very early times. Then Equations \eqref{final eq1}-\eqref{final eq2} becomes

\begin{eqnarray}
\frac{d}{dt}\left(t\Psi_{q}\right)&=&-\frac{1}{t}\delta_{q}\;,\label{rad eq1}\\
\dot{\delta}_{q}&=&-\Psi_{q}\;,\label{rad eq2}
\end{eqnarray}

and

\begin{equation}\label{rad eq3}
\frac{d}{dt}\left(\frac{\delta u_{q}}{\sqrt{t}}\right)=-\frac{1}{t}\delta_{q}\;.
\end{equation}

While equations \eqref{fdelta eq1}-\eqref{fdelta eq2} becomes

\begin{eqnarray}\label{tilde rad eq1}
\left[2\dot{F}\frac{\dot{a}}{a}+\ddot{F}\right]a^2\Psi_q+\left[6F\frac{\dot{a}}{a}+\frac{5}{2}\dot{F}\right]a^2\dot{\Psi}_q+3Fa^2\ddot{\Psi}_q\nonumber\\-\frac{d}{dt}\left(a^2\dot{\tilde{\Psi}}_q\right)&=&\frac{a^2}{t^2}(\tilde{\delta}_q-F\delta_q)\;,\\
\dot{\tilde{\delta}}_q+\dot{\tilde{\Psi}}_q-\partial_0({F\Psi_q})&=&0\;,\\
\frac{\tilde{\delta}_q}{3a}+\frac{d}{dt}\left(\frac{\delta\tilde{u}_q}{a}\right)-F\frac{d}{dt}\left(\frac{\delta u_q}{a}\right)&=&0\;.\label{tilde rad eq2}
\end{eqnarray}

Inspection of Eq. \eqref{F function} show that at this era, for $a\ll C$, we have $F\propto -L_2a\sqrt{C}/3$. Also in DG, time can be integrated from first Friedmann equation with only radiation and matter, one gets:

\begin{equation}\label{time}
t(Y) = \frac{2\sqrt{1+C}}{3H_0}\left(\sqrt{Y+C}(Y-2C) + 2C^{\frac{3}{2}}\right)\;,
\end{equation}

we recall that $Y=a/a_0=a$ assuming $a_0=1$, $H_0=\dot{a}_0/a_0$ is the usual Hubble parameter which we recall is not longer the physical Hubble parameter. Thus, radiation era time and $a(t)$ were related by $a(t)=(3H_0\sqrt{C}/\sqrt{1+C})^{1/2}t^{1/2}$. This complete system Eqs. (\ref{rad eq1})-(\ref{rad eq3}) and Eqs. (\ref{tilde rad eq1})-(\ref{tilde rad eq2}) has analytical solution:

\begin{equation}\label{eins sol rad1}
\delta_{\gamma q}=\delta_{Bq}=\delta_{Dq}=\delta_{\nu q}=\frac{q^{2}t^{2}{\cal R}_{q}}{a^{2}}\;,
\end{equation}

\begin{equation}\label{eins sol rad2}
\dot{\Psi}_{q}=-\frac{tq^{2}{\cal R}_{q}}{a^{2}}\;,
\end{equation}

\begin{equation}\label{eins sol rad3}
\delta u_{\gamma q}=\delta u_{\nu q}=-\frac{2t^{3}q^{2}{\cal R}_{q}}{9a^{2}}\;,
\end{equation}

where\footnote{the definition of ${\cal R}_q$ is given in section 5.4: Conservation outside the horizon, Cosmology, Weinberg.}

\begin{equation}\label{adiabatic norma}
q^{2}{\cal R}_{q}\equiv-a^{2}H\Psi_{q}+4\pi Ga^{2}\delta\rho_{q}+q^{2}H\delta u_{q}\;,
\end{equation}

is a gauge invariant quantity, which take a time independent value for $q/a\ll H$. Here $H=\dot{a}/a$ is the GR definition of the Hubble parameter, which we recall is not longer the physical one. On the other hand, we get

\begin{eqnarray}
\tilde{\delta}_q&=&-\frac{L_2\sqrt{C}q^2{\cal R}_{q}t^2}{3a}\;,\label{delta sol rad1}\\
\dot{\tilde{\Psi}}_q&=&\frac{L_2\sqrt{C}q^2{\cal R}_{q}t}{a}\;,\\
\delta \tilde{u}_q&=&\frac{L_2\sqrt{C}q^2{\cal R}_{q}t^3}{a}\label{delta sol rad3}\,.
\end{eqnarray}

We will talk about this initial conditions later. Note that Eqs. 
(\ref{final fot})$-$ec. (\ref{final bar}) give

\begin{equation}\label{relation fot-bar}
    \frac{d}{dt}\left(\delta_{\gamma}-\delta_{B}\right)=0\;.
\end{equation}

This implies that if we start from adiabatic solutions, $\delta_{\gamma}=\delta_{B}$ is true for all the Universe evolution (the same happens for its delta version, from Eqs. (\ref{final delta fot})$-$Eq (\ref{final delta B})).


\subsection{Matter era}

In this era we use $a\propto t^{2/3}$, then (still using $R=\tilde{R}=0$) we have
\begin{subequations}
\begin{equation}\label{matt era gr1}
\frac{d}{dt}\left(a^{2}\Psi_{q}\right)=-4\pi G\bar{\rho}_{D}a^{2}\delta_{Dq}\,,
\end{equation}

\begin{equation}
\dot{\delta}_{Dq}=-\Psi_{q}\,,
\end{equation}

\begin{equation}
\frac{d}{dt}\left(\frac{(1+R)\delta u_{\gamma q}}{a}\right)=-\frac{1}{3a}\delta_{\gamma q}\,,
\end{equation}

\begin{equation}\label{matt era gr2}
\frac{d}{dt}\left(\frac{\delta u_{\nu q}}{a}\right)=-\frac{1}{3a}\delta_{\nu q}\,.
\end{equation}

For the delta sector,

\begin{eqnarray} 
\left[2\dot{F}\frac{\dot{a}}{a}+\ddot{F}\right]a^2\Psi_q+\left[6F\frac{\dot{a}}{a}+\frac{5}{2}\dot{F}\right]a^2\dot{\Psi}_q+3Fa^2\ddot{\Psi}_q\nonumber,\\-\frac{d}{dt}\left(a^2\dot{\tilde{\Psi}}_q\right)=\frac{2a^2}{3t^2}\left(\tilde{\delta}_{Dq}-F\frac{\delta_{Dq}}{2}\right)\label{matt era DG1}\,,\\
\dot{\tilde{\delta}}_{\gamma q}-\frac{q^2}{a^2}\left(\delta\tilde{u}_{\gamma q}+F\delta u_{\gamma q}\right)+\dot{\tilde{\Psi}}_q-\partial_0(F\Psi_q)&=&0\,,\\
\dot{\tilde{\delta}}_{Dq}+\dot{\tilde{\Psi}}_q-\partial_0(F\Psi_q)&=&0\,,\\
\frac{\tilde{\delta}_{\gamma q}}{3a}+\frac{d}{dt}\left(\frac{\delta\tilde{u}_{\gamma q}}{a}\right)-F\frac{d}{dt}\left(\frac{\delta u_{\gamma q}}{a}\right)&=&0\label{matt era DG2}\,.
\end{eqnarray}
\end{subequations}
Where (in this era),

\begin{equation}\label{matter scale factor}
    a(t)=\left(\frac{3H_0}{2\sqrt{1+C}}\right)^{2/3}t^{2/3},
\end{equation} 

\begin{eqnarray}\label{F func DG matter}
    F(t) &\propto& -\frac{L}{3}a(t)^{3/2}.
\end{eqnarray}

It is remarkable that in GR sector there are exact solutions, given by

\begin{eqnarray}\label{GR pert solutions 1}
  \delta_{Dq}&=&\frac{9q^2t^2{\cal R}_q{\cal T}(\kappa)}{10a^2}\,,\\
  \dot{\Psi}_q&=&-\frac{3q^2t{\cal R}_q{\cal T}(\kappa)}{5a^2}\,,
\end{eqnarray}

\begin{eqnarray}\label{GR pert solutions 2}
  \delta_{\gamma q}=\delta_{\nu q}=\frac{3{\cal R}_q}{5}\left[{\cal T}(\kappa)-{\cal S}(\kappa)\cos\left(q\int_0^t\frac{dt}{\sqrt{3}a}+\Delta(\kappa)\right)\right]\;,\\
  \delta u_{\gamma q}=\delta u_{\nu q}=\frac{3t{\cal R}_q}{5}\left[-{\cal T}(\kappa)+{\cal S}(\kappa)\frac{a}{\sqrt{3}qt}\sin\left(q\int_0^t\frac{dt}{\sqrt{3}a}+\Delta(\kappa)\right)\right]\,.
\end{eqnarray}

Where ${\cal T}(\kappa)$, ${\cal S}(\kappa)$ and $\Delta(\kappa)$ are time-independent dimensionless functions of the dimensionless re-scaled wave number

\begin{equation}\label{kappa}
\kappa \equiv \frac{q\sqrt{2}}{a_{EQ}H_{EQ}}
\end{equation}\,.

$a_{EQ}$ and $H_{EQ}$ are, respectively, the Robertson-Walker scale factor and the expansion rate at matter-radiation equally. These are known as transfer functions. (These functions can only depend on $\kappa$ because they must be independent of the spatial coordinates' normalization and are dimensionless. A complete discussion of the behavior of these functions can be found in \cite{weinberg2008cosmology}). On the other side, delta perturbations have not an exact solution, and numerical calculation is needed to find them, however we will not present numerical solutions in this work, and we only will estimate the initial conditions of the perturbations at the end of this section.

In order to get all transfer functions we have to compare solutions with the full equation system (with $\rho_B=\tilde{\rho}_B=0$). To do this task let us make the change of variable $y\equiv a/a_{EQ}=a/C$, this means

\begin{equation}\label{change of variable}
\frac{d}{dt}=\frac{H_{EQ}}{\sqrt{2}}\frac{\sqrt{1+y}}{y}\frac{d}{dy}\,.
\end{equation}

Also, we will use the following parametrization for all perturbations

\begin{eqnarray*}
  \delta_{Dq}=\kappa^2{\cal R}^0_q d(y)/4\,\,,\;\;\;\;\delta_{\gamma q}=\delta_{\nu q}=\kappa^2{\cal R}^0_qr(y)/4\,,\\
  \dot{\Psi}_q=(\kappa^2H_{EQ}/4\sqrt{2}){\cal R}^0_qf(y)\,\,,\;\;\;\delta u_{\gamma q}=\delta u_{\nu q}=(\kappa^2\sqrt{2}/4H_{EQ}){\cal R}^0_q g(y)\,,
\end{eqnarray*}

and

\begin{eqnarray*}
  \tilde{\delta}_{Dq}=\kappa^2{\cal R}^0_q\tilde{d}(y)/4\,\,,\;\;\;\;\tilde{\delta}_{\gamma q}=\tilde{\delta}_{\nu q}=\kappa^2{\cal R}^0_q\tilde{r}(y)/4\,,\\
  \dot{\tilde{\Psi}}_q=(\kappa^2H_{EQ}/4\sqrt{2}){\cal R}^0_q\tilde{f}(y)\,\,,\;\;\;\delta\tilde{u}_{\gamma q}=\delta\tilde{u}_{\nu q}=(\kappa^2\sqrt{2}/4H_{EQ}){\cal R}^0_q\tilde{g}(y)\,.
\end{eqnarray*}

Then Eqs. \eqref{matt era gr1}-\eqref{matt era gr2} and Eqs. (\ref{matt era DG1})-(\ref{matt era DG2}) become
\begin{subequations}
\begin{eqnarray}
  \sqrt{1+y}\frac{d}{dy}\left(y^2{f}(y)\right)&=&-\frac{3}{2}{d}(y)-\frac{4{r}(y)}{y}\,,\label{Gr eq 1}\\
  \sqrt{1+y}\frac{d}{dy}{r}(y)-\frac{\kappa^2{g}(y)}{y}&=&-y{f}(y)\,,\\
  \sqrt{1+y}\frac{d}{dy}{d}(y)&=&-y{f}(y)\,,\\
  \sqrt{1+y}\frac{d}{dy}\left(\frac{{g}(y)}{y}\right)&=&-\frac{{r}(y)}{3}\,\label{Gr eq 2}\,,
\end{eqnarray}
and
\begin{eqnarray}
-\left[(1+2y)yF'(y)+y(1+y)F''(y)\right]d(y)+\left[6F(y)+\frac{5}{2}yF'(y)\right]y\sqrt{1+y}f(y)&&\nonumber\\+3F(y)y^2\sqrt{1+y}f'(y)-\sqrt{1+y}\frac{d}{dy}\left(y^2\tilde{f}(y)\right)=\frac{3\tilde{d}(y)}{2}+\frac{4\tilde{r}(y)}{y}&&\nonumber\\-\frac{3F(y)d(y)}{4}-\frac{4F(y)r(y)}{y}&&\label{Dg eq 1}\,,\\
\sqrt{1+y}\frac{d}{dy}\tilde{d}(y)=-y\tilde{f}(y)-\sqrt{1+y}\frac{d}{dy}d(y)&&\,,\\
\sqrt{1+y}\frac{d}{dy}\tilde{r}(y)=\frac{\kappa^2}{y}[\tilde{g}(y)+F(y)g(y)]-y\tilde{f}(y)-\sqrt{1+y}\frac{d}{dy}d(y)&&\,,\\
\sqrt{1+y}\frac{d}{dy}\left(\frac{\tilde{g}(y)}{y}\right)=-\frac{\tilde{r}(y)}{3}+\sqrt{1+y}F(y)\frac{d}{dy}\left(\frac{g(y)}{y}\right)&&\label{Dg eq 2}\,.
\end{eqnarray}
\end{subequations}
In this notation, the initial conditions are

\begin{eqnarray*}
  d(y) &=& r(y)\rightarrow y^2\,, \\
  f(y)&\rightarrow& -2\,, \\
  g(y)&\rightarrow& -\frac{y^4}{9}\,.
\end{eqnarray*}

For delta sector,

\begin{eqnarray*}
\tilde{d}(y)&=&\tilde{r}(y)\rightarrow -\frac{L_2C^{3/2}}{3}y^3\,,\\
\tilde{f}(y)&\rightarrow&\sqrt{2}L_2C^{3/2}y\,,\\
\tilde{g}(y)&\rightarrow&\frac{L_2C^{3/2}}{2}y^5\,.
\end{eqnarray*}

From supernovae fit, we know that $C\sim 10^{-4}$ and $L\sim 0.45$ \cite{Alfaro:2012yu,Universe5020051}, thus we can estimate that fluctuations of ``delta matter'' at the beginning of the Universe was much smaller than fluctuations of standard matter. For example, at $y\sim 10^{-3}$ the ratio between components of the Universe is $|\tilde{\delta}_{\alpha}/\delta_{\alpha}|\sim 10^{-10}$.

We do not show numerical solutions here because the aim of this work is to trace a guide for future work, in particular, in the numeric computation of multipole coefficients for temperature fluctuations in the CMB. However, we will derive the equations to do that computation.

\section{Derivation of temperature fluctuations}\label{sec:fluctuation temperature}

It is possible to find expressions analogous to temperature fluctuations usually obtained by Boltzmann equations by studying photons propagation in FRLW perturbed coordinates, with the condition $\bar{g}_{i0}=0$\footnote{see Section 7.1: General formulas for the temperature fluctuation, Cosmology, Weinberg.}. For DG, the metric which photons follow is given by

\begin{eqnarray}\label{metric}
 {\bf g}_{00}&=&-((1+3F(t))c^2+E({\bf x},t)+\tilde{E}({\bf x},t))\,,\;\; \;\;{\bf g}_{i0}=0\,,\nonumber\\
 {\bf g}_{ij}&=&a^2(t)(1+F(t))\delta_{ij}+h_{ij}({\bf x},t)+\tilde{h}_{ij}({\bf x},t)\,,
\end{eqnarray}

A ray of light propagating to the origin of the FRLW coordinate system , from a direction $\hat{n}$, will have a comoving radial coordinate $r$ related with $t$ by

\begin{eqnarray}\label{line element}
  0={\bf \bar{g}_{\mu\nu}}dx^{\mu}dx^{\nu}=-((1+3F(t))c^2+E(r\hat{n},t)+\tilde{E}(r\hat{n},t))dt^2\nonumber\\+(a^2(t)(1+F(t))+h_{rr}(r\hat{n},t)+\tilde{h}_{rr}(r\hat{n},t))dr^2\,,
\end{eqnarray}

in other words,

\begin{eqnarray}\label{dr/dt}
\frac{dr}{dt}&=&-\left(\frac{(1+3F(t))c^2+E+\tilde{E}}{a^2(t)(1+F(t))+h_{rr}+\tilde{h}_{rr}}\right)^{1/2}\nonumber\\&\simeq& -\frac{c}{a_{DG}(t)}+\frac{c(h_{rr}+\tilde{h}_{rr})}{2(1+3F(t))a_{DG}^3(t)}-\frac{E+\tilde{E}}{2(1+3F(t))ca_{DG}(t)}\,,
\end{eqnarray}

where $a_{DG}(t)$ is the modified scale factor given by

\begin{equation}\label{msf}
  a_{DG}(t)=a(t)\sqrt{\frac{1+F(t)}{1+3F(t)}}.
\end{equation}

Now we will use the approximation of a sharp transition between opaque and transparent Universe at a moment $t_{ls}$ of last scattering, at red shift $z\simeq 1090$. With this approximation, the relevant term at first order in Eq. (\ref{dr/dt}) is

\begin{equation}\label{r(t)}
r(t)=c\left[s(t)+\int_{t_{ls}}^t\frac{dt'}{a_{DG}(t')}N\left(cs(t')\hat{n},t'\right)\,\right]\,,
\end{equation}

where

\begin{equation}\label{perturbation}
N({\bf x},t)\equiv \frac{1}{2(1+3F)}\left[\frac{h_{rr}({\bf x},t)+\tilde{h}_{rr}({\bf x},t)}{a_{DG}^2}-\frac{E({\bf x},t)}{c^2}-\frac{\tilde{E}({\bf x},t)}{c^2}\right]\,,
\end{equation}

and $s(t)$ is the zero order solution for the radial coordinate. $s(t) = r_{ls}$ when $t=t_{ls}$:

\begin{equation}\label{rczo}
s(t)=r_{ls}-\int_{t_{ls}}^t\frac{dt'}{a_{DG}(t')}=\int_t^{t_0}\frac{dt'}{a_{DG}(t')}\,.
\end{equation}

If a ray of light arrives to $r=0$ at a time $t_0$, then Eq. \eqref{r(t)} gives

\begin{equation}\label{r=0}
0=s(t_0)+\int_{t_{ls}}^t\frac{dt'}{a_{DG}(t')}N\left(cs(t')\hat{n},t'\right)=r_{ls}+\int_{t_{ls}}^{t_0}\frac{dt}{a_{DG}(t)}\left(N\left(cs(t)\hat{n},t\right)-1\right)\,.
\end{equation}

A time interval $\delta t_{ls}$, between departure of successive rays of light at time $t_{ls}$ of last scattering, produces an interval of time $\delta t_0$, between the arrival of the rays of light at $t_0$, given by the variation of Eq. \eqref{r=0}:

\begin{eqnarray}\label{variationR}
  0=\frac{\delta t_{ls}}{a_{DG}(t_{ls})}\left[1-N(cr_{ls}\hat{n},t_{ls})+c\int_{t_{ls}}^{t_0}\frac{dt}{a_{DG}(t)}\left(\frac{\partial N\left(r(t)\hat{n},t\right)}{\partial r}\right)_{r=cs(t)}\right]\nonumber\\
  +\delta t_{ls}(\partial u_{\gamma}^{r}(cr_{ls}\hat{n},t_{ls})+\partial \tilde{u}_{\gamma}^r(cr_{ls}\hat{n},t_{ls}))+\frac{\delta t_0}{a_{DG}(t_0)}\left[-1+N(0,t_0)\right]\,.
\end{eqnarray}

The velocity terms of the photon-gas or photon-electron-nucleon arise because of the variation respect to the time of the radial coordinate $r_{ls}$ described by the Eq. \eqref{r=0}. The exchange rate of $N(s(t)\hat{n},t)$ is

\begin{equation}
\frac{d}{dt}N\left(s(t)\hat{n},t\right)=\left(\frac{\partial}{\partial t}N(r\hat{n},t)\right)_{r=cs(t)}-\frac{c}{a_{DG}(t)}\left(\frac{\partial N(r\hat{n},t)}{\partial r}\right)_{r=cs(t)}\,,\nonumber
\end{equation}

then, 

\begin{eqnarray}\label{variation}
0=\frac{\delta t_{ls}}{a_{DG}(t_{ls})}\left[1-N(0,t_{ls})+\int_{t_{ls}}^{t_0}dt\left(\frac{\partial N\left(r\hat{n},t\right)}{\partial t}\right)_{r=cs(t)}\right]\nonumber\\
  +\delta t_{ls}(\partial u_{\gamma}^{r}(r_{ls}\hat{n},t_{ls})+\partial \tilde{u}_{\gamma}^r(r_{ls}\hat{n},t_{ls}))+\frac{\delta t_0}{a_{DG}(t_0)}\left[-1+N(0,t_0)\right]\,.
\end{eqnarray}

This result gives the ratio between the time intervals between ray of lights that are emitted and received. However, we are interested in this ratio, but for the proper time, that in DG it is defined with the original metric $g_{\mu\nu}$:

\begin{equation}\label{proper time}
\delta \tau_L=\sqrt{1+\frac{E(r_{ls},t_{ls})}{c^2}}\delta t_{ls}\;,\;\;\;\delta\tau_0=\sqrt{1+\frac{E(0,t_0)}{c^2}}\delta t_{0}\;,
\end{equation}

At first order, it gives the ratio between a received frequency and an emitted one: 

\begin{eqnarray}\label{ratio of frequencies}
  \frac{\nu_0}{\nu_L}=\frac{\delta \tau_{L}}{\delta \tau_{0}}=\frac{a_{DG}(t_{ls})}{a_{DG}(t_0)}\left[1+\frac{1}{2c^2}\left(E(r_{ls}\hat{n},t)-E(0,t_0)\right)\right.\nonumber\\-\left.\int_{t_{ls}}^{t_0}\left(\frac{\partial}{\partial t}N(r\hat{n},t)\right)_{r=cs(t)}dt-a_{DG}(t)(\delta u_{\gamma}^{r}(r_{ls}\hat{n},t)+\delta \tilde{u}_{\gamma}^r(r_{ls}\hat{n},t))\right]\,.
\end{eqnarray}

In \cite{Universe5020051}, we defined the physical scale factor as $Y_{DG}(t)\equiv a_{DG}(t)/a_{DG}(t_0)$. Thus, we recover the standard expression for the redshift. The observed temperature at the present time $t_0$ from direction $\hat{n}$ is

\begin{equation}\label{temp fluct}
T(\hat{n})=\left(\frac{\nu_0}{\nu_L}\right)(\bar{T}(t_{ls})+\delta T(cr_{ls}\hat{n},t_{ls}))\;,
\end{equation}

In absence of perturbations, the observed temperature in all directions should be

\begin{equation}\label{temp}
T_0=\left(\frac{a_{DG}(t_{ls})}{a_{DG}(t_0)}\right)\bar{T}(t_{ls})\,,
\end{equation}

therefore, the ratio between the observed temperature shift that comes from direction $\hat{n}$ and the unperturbed value is

\begin{eqnarray}\label{ratio temp}
  \frac{\Delta T(\hat{n})}{T_0}&\equiv& \frac{T(\hat{n})-T_0}{T_0}=\frac{\nu_0a_{DG}(t_0)}{\nu_La_{DG}(t_{ls})}-1+\frac{\delta T(cr_{ls}\hat{n},t_{ls})}{\bar{T}(t_{ls})}\nonumber\\
  &=&\frac{1}{2c^2}\left(E(r_{ls}\hat{n},t)-E(0,t_0)\right)-\int_{t_{ls}}^{t_0}dt\left(\frac{\partial}{\partial t}N(r\hat{n},t)\right)_{r=cs(t)}\nonumber\\&-&a_{DG}(t)(\delta u_{\gamma}^{r}(r_{ls}\hat{n},t)+\delta \tilde{u}_{\gamma}^r(r_{ls}\hat{n},t))+\frac{\delta T(cr_{ls}\hat{n},t_{ls})}{\bar{T}(t_{ls})}\,.
\end{eqnarray}

For scalar perturbations in any gauge with ${\bf h}_{i0}=0$, the metric perturbations are

\begin{eqnarray}\label{metric pert}
  h_{00}=-E\;\;,\;\;h_{ij}=(1+F)a^2\left[A\delta_{ij}+\frac{\partial^2B}{\partial x^i\partial x^j}\right], \nonumber\\
  \tilde{h}_{00}=-\tilde{E}\;\;,\;\;\tilde{h}_{ij}=(1+F)a^2\left[\tilde{A}\delta_{ij}+\frac{\partial^2\tilde{B}}{\partial x^i\partial x^j}\right].
\end{eqnarray}

Besides for scalar perturbations radial velocity of the photon fluid and the delta versions are given in terms of the velocity potentials  $\delta u_{\gamma}$ and $\delta \tilde{u}_{\gamma}$, respectively,

\begin{eqnarray}\label{vel pert}
  \delta u_{\gamma}^r=(\bar{g}+\bar{\tilde{g}})^{r\mu}\frac{\partial \delta u_{\gamma}}{\partial x^{\mu}}=\frac{1}{(1+ F(t))a^2}\frac{\partial \delta u_{\gamma}}{\partial r}\,,\nonumber\\
  \delta \tilde{u}_{\gamma}^r=(\bar{g}+ \bar{\tilde{g}})^{r\mu}\frac{\partial \delta \tilde{u}_{\gamma}}{\partial x^{\mu}}=\frac{1}{(1+ F(t))a^2}\frac{\partial \delta \tilde{u}_{\gamma}}{\partial r}\,.
\end{eqnarray}

Then Eq. \eqref{ratio temp} gives the scalar contribution to temperature fluctuations

\begin{eqnarray}\label{scalar ratio temp}
  \left(\frac{\Delta T(\hat{n})}{T_0}\right)^S&=&\frac{1}{2c^2}\left(E(r_{ls}\hat{n},t)-E(0,t_0)\right)-\int_{t_{ls}}^{t_0}dt\left(\frac{\partial}{\partial t}N(r\hat{n},t)\right)_{r=cs(t)}\nonumber\\&-&\frac{1}{(1+3F(t))a_{DG}}\left(\frac{\partial \delta u_{\gamma}(cr_{ls}\hat{n},t)}{\partial r}+\frac{\partial \delta \tilde{u}_{\gamma}(cr_{ls}\hat{n},t)}{\partial t}\right)\nonumber\\&+&\frac{\delta T(cr_{ls}\hat{n},t_{ls})}{\bar{T}(t_{ls})}\,,
\end{eqnarray}

where

\begin{equation}\label{N pert}
N=\frac{1}{2}\left[A+\frac{\partial^2B}{\partial r^2}+\left(\tilde{A}+\frac{\partial^2\tilde{B}}{\partial r^2}\right)-\frac{E}{1+3 F}-\frac{\tilde{E}}{1+3F}\right]\,.
\end{equation}

In the next step we will study the gauge transformations of these fluctuations. The following identity for the fields $B$ and $\tilde{B}$ will be useful:

\begin{equation}
  \left(\frac{\partial^2\dot{B}}{\partial r^2}\right)_{r=s(t)}=-\left(\frac{d}{dt}\left[a_{DG}\frac{\partial \dot{B}}{\partial r}+a_{DG}\dot{a}_{DG}\dot{B}+a_{DG}^2\ddot{B}\right]+\frac{\partial}{\partial t}\left[a_{DG}\dot{a}_{DG}\dot{B}+a_{DG}^2\ddot{B}\right]\right)_{r=s(t)}.
\end{equation}

Then, the temperature fluctuations are described by

\begin{equation}
  \left(\frac{\Delta T(\hat{n})}{T_0}\right)^S=\left(\frac{\Delta T(\hat{n})}{T_0}\right)^S_{early}+\left(\frac{\Delta T(\hat{n})}{T_0}\right)^S_{late}+\left(\frac{\Delta T(\hat{n})}{T_0}\right)^S_{ISW}
\end{equation}

where

\begin{eqnarray}\label{early}
  \left(\frac{\Delta T(\hat{n})}{T_0}\right)^S_{early}&=&-\frac{1}{2}a_{DG}(t_{ls})\dot{a}_{DG}(t_{ls})\dot{B}(r_{ls}\hat{n},t_{ls})-\frac{1}{2}a_{DG}^2(t_{ls})\ddot{B}(r_{ls}\hat{n},t_{ls})
  +\frac{1}{2}E(r_{ls}\hat{n},t_{ls})\nonumber\\+\frac{\delta T(r_{ls}\hat{n})}{\bar{T}(t_{ls})}
  &-&a_{DG}(t_{ls})\left[\frac{\partial}{\partial r}\left(\frac{1}{2}\dot{B}(r\hat{n},t_{ls})+\frac{1}{(1+3F(t_{ls}))a_{DG}^2(t_{ls})}\delta{u}_{\gamma}(r\hat{n},t_{ls})\right)_{r=r_{ls}}\right]\nonumber\\
  &-&\left\{\left(\frac{1}{2}a_{DG}(t_{ls})\dot{a}_{DG}(t_{ls})\dot{\tilde{B}}(r_{ls}\hat{n},t_{ls})+\frac{1}{2}a_{DG}^2(t_{ls})\ddot{\tilde{B}}(r_{ls}\hat{n},t_{ls})\right)\right.\nonumber\\+a_{DG}(t_{ls})&\times&\left.\left[\frac{\partial}{\partial r}\left(\frac{1}{2}\dot{\tilde{B}}(r\hat{n},t_{ls})+\frac{1}{(1+3F(t_{ls}))a_{DG}^2(t_{ls})}\delta{\tilde{u}}_{\gamma}(r\hat{n},t_{ls})\right)_{r=r_{ls}}\right]\right\}\,,
\end{eqnarray}

\begin{eqnarray}\label{late}
  \left(\frac{\Delta T(\hat{n})}{T_0}\right)^S_{late}&=&\frac{1}{2}a_{DG}(t_0)\dot{a}_{DG}(t_0)\dot{B}(0,t_0)+\frac{1}{2}a_{DG}^2(t_0)\ddot{B}(0,t_0)-\frac{1}{2}E(0,t_0)\nonumber\\
  &+&a_{DG}(t_0)\left[\frac{\partial}{\partial r}\left(\frac{1}{2}\dot{B}(r\hat{n},t_0)+\frac{1}{(1+3F(t_0))a_{DG}^2(t_0)}\delta{u}_{\gamma}(r\hat{n},t_0)\right)_{r=0}\right]\nonumber\\
  &+&\left\{\left(\frac{1}{2}a_{DG}(t_0)\dot{a}_{DG}(t_0)\dot{\tilde{B}}(0,t_0)+\frac{1}{2}a_{DG}^2(t_0)\ddot{\tilde{B}}(0,t_0)\right)\right.\nonumber\\+a_{DG}(t_0)&\times&\left.\left[\frac{\partial}{\partial r}\left(\frac{1}{2}\dot{\tilde{B}}(r\hat{n},t_0)+\frac{1}{(1+3F(t_0))a_{DG}^2(t_0)}\delta{\tilde{u}}_{\gamma}(r\hat{n},t_0)\right)_{r=r_{ls}}\right]\right\}\;,
\end{eqnarray}

\begin{eqnarray}\label{ISW}
  \left(\frac{\Delta T(\hat{n})}{T_0}\right)^S_{ISW}&=&-\frac{1}{2}\int_{t_{ls}}^{t_0}dt\left\{\frac{\partial}{\partial t}\left[a_{DG}^2(t)\ddot{B}(r\hat{n},t)+a_{DG}(t)\dot{a}_{DG}(t)\dot{B}(r\hat{n},t)+A(r\hat{n},t)\right.\right.\nonumber\\&-&\frac{E(r\hat{n},t)}{1+3 F(t)}
  + a_{DG}^2(t)\ddot{\tilde{B}}(r\hat{n},t)+a_{DG}(t)\dot{a}_{DG}(t)\dot{\tilde{B}}(r\hat{n},t)\nonumber\\&+&\left.\left.\tilde{A}(r\hat{n},t)-\frac{\tilde{E}(r\hat{n},t)}{1+3 F(t)}\right]\right\}\,,
\end{eqnarray}

The ``late'' term is the sum of independent direction terms and a term proportional to $\hat{n}$, which was added to represent the local anisotropies of the gravitational field and the local fluid. In GR, these terms only contribute to the multipole expansion for $l=0$ and $l=1$. Thus we will ignore their contribution to DG.

\subsection{Gauge transformations}

We are going to study the gauge transformations for photons propagating in the metric ${\bf g}_{\mu\nu}$ for a parameter ${\boldsymbol{\epsilon}_{\mu}}$. Then the transformations are

\begin{eqnarray}\label{gauge transtotal}
\Delta {A}&=&\frac{2\dot{a}}{(1+F)a}\frac{\boldsymbol{\epsilon_0}}{1+3 F}\,,\;\;\;\;\Delta{B}=-\frac{2}{1+ F}\frac{\boldsymbol{\epsilon^S}}{(1+ F)a^2}\,,\nonumber\\ \Delta {C}_i&=&-\frac{1}{1+F}\frac{\boldsymbol{\epsilon_i^V}}{(1+ F)a^2}\,,\;\;\;\;\Delta {D}_{ij}=0\,,\;\;\;\;\Delta {E}=2\frac{\partial}{\partial t}\left(\frac{\boldsymbol{\epsilon_0}}{1+3 F}\right),\\ \Delta {H}&=&-\frac{1}{\sqrt{1+F}a}\left[a^2\frac{\partial}{\partial t}\left(\frac{\boldsymbol{\epsilon^S}}{(1+F)a^2}\right)+\frac{\boldsymbol{\epsilon_0}}{(1+3 F)}\right]\,,\;\;\;\;\Delta {G}_i=-\frac{a}{\sqrt{1+ F}}\frac{\partial }{\partial t}\left(\frac{\boldsymbol{\epsilon_i^V}}{(1+ F)a^2}\right).\nonumber
\end{eqnarray}

and

\begin{eqnarray}\label{delta gauge trans}
\Delta \tilde{A}&=&\frac{1}{(1+F)a^2}\left[\frac{\partial}{\partial t}(Fa^2)\frac{\boldsymbol{\epsilon_0}}{1+3 F}\right]\,,\;\;\;\;\Delta\tilde{B}=-\frac{1}{(1+F)a^2}\left[\frac{2F}{1+F}\boldsymbol{\epsilon^S}\right]\,,\nonumber\\ \Delta \tilde{C}_i&=&-\frac{F}{1+F}\frac{\boldsymbol{\epsilon_i^V}}{(1+F)a^2}\,,\;\;\;\;\Delta \tilde{D}_{ij}=0\,,\;\;\;\;\Delta \tilde{E}=6F\frac{\partial}{\partial t}\left(\frac{\boldsymbol{\epsilon_0}}{1+3F}\right)+\frac{3\dot{F}}{1+3 F}\boldsymbol{\epsilon_0}\nonumber\\ \Delta \tilde{H}&=&-\frac{1}{\sqrt{1+F}a}\left[Fa^2\frac{\partial}{\partial t}\left(\frac{\boldsymbol{\epsilon^S}}{(1+F)a^2}\right)+\frac{3F\boldsymbol{\epsilon_0}}{(1+3 F)}\right]\,,\nonumber\\\Delta \tilde{G}_i&=&-\frac{1}{\sqrt{1+F}a}\left[Fa^2\frac{\partial }{\partial t}\left(\frac{\boldsymbol{\epsilon_i^V}}{(1+ F)a^2}\right)\right].
\end{eqnarray}

Now, considering the sum of the perturbations we get
\begin{subequations}
\begin{eqnarray}
    \Delta A+\Delta\tilde{A}&=&\frac{1}{(1+F)a^2}\frac{\partial }{\partial t}[(1+F)a^2]\frac{\boldsymbol{\epsilon_0}}{1+3F}\,,\\
    \Delta B+\Delta\tilde{B}&=&-\frac{2\boldsymbol{\epsilon^S}}{(1+F)a^2}\,,\\
    \Delta E+\Delta\tilde{E}&=&2(1+3F)\frac{\partial}{\partial t}\left(\frac{\boldsymbol{\epsilon_0}}{1+3F}\right)+\frac{3\dot{F}}{1+3F}\boldsymbol{\epsilon_0}\,,\\
    \Delta H+\Delta\tilde{H}&=&-\frac{1}{\sqrt{1+F}a}\left[(1+F)a^2\frac{\partial}{\partial t}\left(\frac{\boldsymbol{\epsilon^S}}{(1+F)a^2}\right)+\boldsymbol{\epsilon_0}\right]\,,\\
    \Delta C_i+\Delta\tilde{C}_i&=&-\frac{\boldsymbol{\epsilon_i^V}}{(1+F)a^2}\,,\\
    \Delta G_i+\Delta\tilde{G}_i&=&-\frac{1}{\sqrt{1+F}a}\left[(1+F)a^2\frac{\partial }{\partial t}\left(\frac{\boldsymbol{\epsilon_i^V}}{(1+F)a^2}\right)\right]\,.
\end{eqnarray}
\end{subequations}

Now, we will study the gauge transformations that preserve the condition ${\bf g}_{i0}=g_{io}+\tilde{g}_{i0}=0$. This means that $\Delta H+\Delta \tilde{H}=0$. This gives a solution for $\boldsymbol{\epsilon_0}$ given by

\begin{eqnarray}
  \boldsymbol{\epsilon_0}=-(1+F)a^2\frac{\partial}{\partial t}\left(\frac{\boldsymbol{\epsilon^S}}{(1+F)a^2}\right)\,.
\end{eqnarray}

When we study how ``ISW'' term transform under this type of transformations, we found that $\Delta ISW=0$. While for the ``early'' term we should note that temperature perturbations transforms as

\begin{equation}
  \Delta \delta T(r_{ls}\hat{n},t)=\dot{\bar{T}}(t)\frac{\boldsymbol{\epsilon_0}}{1+3F}\,,
\end{equation}

With this expression and $\bar{T}a_{DG}=cte$, we finally obtain

\begin{equation}
  \frac{\Delta \delta T(r_{ls}\hat{n},t)}{\bar{T}(t_{ls})}=-\frac{\dot{a}_{DG}}{a_{DG}}\frac{\boldsymbol{\epsilon_0}}{1+3F}\,.
\end{equation}

This results implies that the ``early'' term is invariant under this gauge transformation. Note that this gauge transformation is equivalent to the previously discussed in Section 1, because we can always take $\boldsymbol{\epsilon}$ as a combination of $\epsilon$ and $\tilde{\epsilon}$. Then we remark that temperature fluctuations are gauge invariant under scalar transformations that leave ${\bf g}_{i0}=0$.

\subsection{Single modes}

We will assume that since the last scattering until now all the scalar contributions are dominated by a unique mode, such that any perturbation $X({\bf x},t)$ could be written as

\begin{equation}
  X({\bf x},t)=\int d^3q\alpha({\bf q})e^{i{\bf q\cdot x}}X_q(t)\,,
\end{equation}

where $\alpha({\bf q})$ is an stochastic variable, normalized such that

\begin{equation}
\langle\alpha({\bf q})\alpha^*({\bf q}')\rangle=\delta^3({\bf q}-{\bf q}')\,.
\end{equation}

Then Eqs \eqref{early} and \eqref{ISW} become

\begin{eqnarray}\label{flucff}
  \left(\frac{\Delta T(\hat{n})}{T_0}\right)^S_{early}&=&\int d^3q \alpha({\bf q})e^{i{\bf q}\cdot\hat{n}r(t_{ls})}\left({\cal F}(q)+\tilde{{\cal F}}(q)+i\hat{q}\cdot\hat{n}({\cal G}(q)+\tilde{{\cal G}}(q))\right)\,,\\
  \left(\frac{\Delta T(\hat{n})}{T_0}\right)^S_{ISW}&=&-\frac{1}{2}\int_{t_0}^{t_1}dt\int d^3q\alpha({\bf q})e^{i{\bf q}\cdot\hat{n}s(t)}\frac{d}{dt}\left[a_{DG}^2(t)\ddot{B}_q(t)+a_{DG}(t)\dot{a}_{DG}(t)\dot{B}_q(t)\right.\nonumber\\&+&A_q(t)-\frac{E_q(t)}{1+3F(t)}
 +\left.\left(a_{DG}^2(t)\ddot{\tilde{B}}_q(t)+a_{DG}(t)\dot{a}_{DG}(t)\dot{\tilde{B}}_q(t)+\tilde{A}_q(t)\right.\right.\nonumber\\&-&\left.\left.\frac{\tilde{E}_q(t)}{1+3 F(t)}\right)\right]\,,
\end{eqnarray}

where

\begin{eqnarray}
  {\cal F}(q)&=&-\frac{1}{2}a_{DG}^2(t)\ddot{B}_q(t_{ls})-\frac{1}{2}a_{DG}(t)\dot{a}_{DG}(t_{ls})\dot{B}_q(t_{ls})+\frac{1}{2}E_q(t_{ls})+\frac{\delta T_q(t_{ls})}{\bar{T}(t_{ls})}\,,\\
  \tilde{{\cal F}}(q)&=&-\frac{1}{2}a_{DG}^2(t)\ddot{\tilde{B}}_q(t_{ls})-\frac{1}{2}a_{DG}(t_{ls})\dot{a}_{DG}(t_{ls})\dot{\tilde{B}}_q(t_{ls})\,,\\
  {\cal G}(q)&=&-q\left(\frac{1}{2}a_{DG}(t_{ls})\dot{B}_q(t_{ls})+\frac{1}{(1+3F(t_{ls}))a_{DG}(t_{ls})}\delta u_{\gamma}(t_{ls})\right)\,,\\
  \tilde{{\cal G}}(q)&=&-q\left(\frac{1}{2}a_{DG}(t_{ls})\dot{\tilde{B}}_q(t_{ls})+\frac{1}{(1+3F(t_{ls}))a_{DG}(t_{ls})}\delta \tilde{u}_{\gamma}(t_{ls})\right)\,.
\end{eqnarray}

These functions are called form factors. We emphasize that combination given by ${\cal F}(q)+\tilde{{\cal F}}(q)$ and ${\cal G}(q)+\tilde{{\cal G}}(q)$, and the expression inside the integral are gauge invariants under gauge transformations that preserve ${\bf g}_{i0}$ equal to zero.

\section{Coefficients of multipole temperature expansion: Scalar modes}

As an application of the previous results, we will study the contribution of the scalar modes for temperature-temperature correlation, given by:

\begin{equation}\label{coeff T-T}
C_{TT,l}=\frac{1}{4\pi}\int d^2\hat{n}\int d^2\hat{n}'P_{l}(\hat{n}\cdot\hat{n}')\langle\Delta T(\hat{n})\Delta T(\hat{n}')\rangle\,,
\end{equation}

where $\Delta T(\hat{n})$ is the stochastic variable which gives the deviation of the average of observed temperature in direction $\hat{n}$, and $\langle \ldots\rangle$ denotes the average over the position of the observer. However, the observed quantity is

\begin{equation}
C_{TT,l}^{obs}=\frac{1}{4\pi}\int d^2\hat{n}\int d^2\hat{n}'P_l(\hat{n}\cdot\hat{n}')\Delta T(\hat{n})\Delta T(\hat{n'})\,,
\end{equation}

nevertheless, the mean square fractional difference between this equation and Eq. \eqref{coeff T-T} is $2/(2l+1)$, and therefore it may be neglected for $l\gg 1$.\\
In order to calculate this coefficients we use the following expansion in spherical harmonics

\begin{equation}
  e^{i\hat{q}\cdot\hat{n}\rho}=4\pi\sum_{l=0}^{\infty}\sum_{m=-l}^{m=l}i^lj_l(\rho)Y_l^m(\hat{n})Y_{l}^{m*}(\hat{q})\,,
\end{equation}

where $j_l(\rho)$ are the spherical Bessel's functions. Using this expression in Eq. \eqref{flucff}, and replacing the factor $i\hat{q}\cdot\hat{n}$ for time derivatives of Bessel's functions, the scalar contribution of the observed T-T fluctuations in direction $\hat{n}$ are

\begin{equation}
(\Delta T(\hat{n}))^S=\sum_{lm}a_{T,lm}^SY_l^m(\hat{n})\,,
\end{equation}

where

\begin{eqnarray}
  a_{T,lm}^S=4\pi i^lT_0\int d^3q\alpha({\bf q})Y_{l}^{l*}(\hat{q})\left[j_l(qr_{ls})({\cal F}(q)+\tilde{{\cal F}}(q))+j_l'(qr_{ls})({\cal G}(q)+\tilde{{\cal G}}(q))\right]\,,
\end{eqnarray}

and $\alpha({\bf q})$ is a stochastic parameter for the dominant scalar mode. It is normalized such that

\begin{equation}
\langle\alpha({\bf q})\alpha^*({\bf q'})\rangle=\delta^3({\bf q}-{\bf q'})\,.
\end{equation}

Inserting this expression in Eq. \eqref{coeff T-T} we get

\begin{equation}\label{coeff1}
  C_{TT,l}^S=16\pi^2T_0^2\int_0^{\infty}q^2dq\left[j_l(qr_{ls})({\cal F}(q)+\tilde{{\cal F}}(q))+j_l'(qr_{ls})({\cal G}(q)+\tilde{{\cal G}}(q))\right]^2\,.
\end{equation}

Now we will consider the case $l\gg 1$. In this limit we can use the following approximation for Bessel's functions\footnote{See, e.g. I. S. Gradsteyn \& I. M. Ryzhik, {\it Table of Integral, Series, and Products}, translated, corrected and enlarged by A. Jeffrey (Academic Press, New York, 1980): formula 8.453.1.}:

\begin{eqnarray}
  j_l(\rho)\rightarrow\left\{\begin{array}{cc} \cos(b)\cos\left[\nu(\tan b-b)-\pi/4\right]/(\nu\sqrt{\sin b})& \rho>\nu\,,\\
  0 & \rho<\nu\,,\end{array}\right.
\end{eqnarray}

where $\nu\equiv l+1/2$, and $\cos b\equiv \nu/\rho$, with $0\leq b\leq \pi/2$. Besides, for $\rho>\nu\gg 1$ the phase $\nu(\tan b -b)$ is a function of $\rho$ that grows very fast, then the derivatives of Bessel's functions only acts in its phase:

\begin{eqnarray}
  j'_l(\rho)\rightarrow\left\{\begin{array}{cc} -\cos(b)\sqrt{\sin b}\sin\left[\nu(\tan b-b)-\pi/4\right]/\nu& \rho>\nu\,,\\
  0 & \rho<\nu\,.\end{array}\right.
\end{eqnarray}

Using these approximations in Eq. \eqref{coeff1} and changing the variable from $q$ to $b=\cos^{-1}(\nu/qr_{ls})$,  we obtain

\begin{eqnarray}\label{coeff2}
  C_{TT,l}^{S}&=&\frac{16\pi^2T_0^2\nu}{r_{ls}^3}\int_0^{\pi/2}\frac{db}{\cos^2b}\nonumber\\
 && \times\left[\left({\cal F}\left(\frac{\nu}{r_{ls}\cos b}\right)+\tilde{{\cal F}}\left(\frac{\nu}{r_{ls}\cos b}\right)\right)\cos[\nu(\tan b -b)-\pi/4]\right.\nonumber\\
  &&  -\left.\sin b \left({\cal G}\left(\frac{\nu}{r_{ls}\cos b}\right)+\tilde{{\cal G}}\left(\frac{\nu}{r_{ls}\cos b}\right)\right)\sin[\nu(\tan b-b)-\pi/4]\right]^2\,.
\end{eqnarray}

When $\nu\gg 1$, the functions $\cos[\nu(\tan b -b)-\pi/4]$ and $\sin[\nu(\tan b -b)-\pi/4]$ oscillate very rapidly, then the squared average of its values are $1/2$, while the averaged cross terms are zero. Using $l\approx \nu$, and changing the integration variable from $b$ to $\beta=1/\cos b$, the Eq. \eqref{coeff2} becomes

\begin{eqnarray}\label{coeff3}
  l(l+1)C_{TT,l}^S&=&\frac{8\pi^2T_0^2l^3}{r_{ls}^3}\int_1^{\infty}\frac{\beta d\beta}{\sqrt{\beta^2-1}}\nonumber\\
  &&\times\left[\left({\cal F}\left(\frac{l\beta}{r_{ls}}\right)+\tilde{{\cal F}}\left(\frac{l\beta}{r_{ls}}\right)\right)^2+\frac{\beta^2-1}{\beta^2}\left({\cal G}\left(\frac{l\beta}{r_{ls}}\right)+\tilde{{\cal G}}\left(\frac{l\beta}{r_{ls}}\right)\right)^2\right]\,.
\end{eqnarray}

Note that $d_A=r_{ls}\tilde{R}_{ls}$ is the angular diameter distance of the last scattering surface. To calculate the CMB power spectrum, we need to know the value of $\dot{\tilde{B}}_q$. We use the off diagonal equation from Delta sector to obtain it. This gives:

\begin{equation}
  \dot{\tilde{A}}_q=\dot{A}_qF+A_q\dot{F}-2a^2(\rho+p)\delta u_q- a^2(\tilde{\rho}+\tilde{p})\delta u_q-(\rho+p)\delta \tilde{u}_q\,,
\end{equation}

so if we use this equation with the definition of $\tilde{\Psi}$

\begin{equation}
\dot{\tilde{\Psi}}_q=\frac{1}{2}(3\dot{\tilde{A}}_q-q^2\dot{\tilde{B}}_q),
\end{equation}

it allow us to find $\dot{\tilde{B}}$. Now we will use the approximation of that perturbations of gravitation field are dominated by perturbations of dark matter density. In this regime $\dot{A}_q(t_{ls})=0$ and in the synchronous gauge, the velocity perturbations for Dark matter are zero, then

\begin{equation}
  \dot{\tilde{A}}_q(t_{ls})=A_q(t_{ls})\dot{F}(t_{ls})\;,
\end{equation}

and

\begin{equation}
  \dot{\tilde{B}}_q(t_{ls})=\frac{3}{q^2}A_q(t_{ls})\dot{F}(t_{ls})-\frac{2\dot{\tilde{\Psi}}_q(t_{ls})}{q^2}\Rightarrow \ddot{\tilde{B}}_q(t_{ls})=\frac{3}{q^2}A_q(t_{ls})\ddot{F}(t_{ls})-\frac{2\ddot{\tilde{\Psi}}_q(t_{ls})}{q^2}\,,
\end{equation}

where

\begin{eqnarray}
  q^2A_q&=&8\pi Ga^2\delta\rho_{Dq}-2Ha^2\dot{\Psi}_q\nonumber\\
  &=&3H^2a^2\delta_{Dq}-2Ha^2\dot{\Psi}_q\;.
\end{eqnarray}

In GR $\dot{B}_q=-2\dot{\Psi}_q/q^2$, and $\dot{\Psi}_q\propto t^{-1/3}$ implies $\ddot{B}_q=2\dot{\Psi}_q/3tq^2$. Therefore,the usual form factors are:

\begin{eqnarray}\label{form factors}
  {\cal F}(q)&=&\frac{1}{3}\delta_{\gamma q}(t_{ls})+\frac{\dot{\Psi}_q(t_{ls})}{q^2}\left(a_{DG}(t_{ls})\dot{a}_{DG}(t_{ls})-\frac{2}{3}\frac{a_{DG}^2(t_{ls})}{t_{ls}}\right)\,,\\
  {\cal G}(q)&=&-q\frac{\delta u_{\gamma q}(t_{ls})}{(1+3F(t_{ls}))a_{DG}(t_{ls})}+\frac{a_{DG}(t_{ls})\dot{\Psi}_q(t_{ls})}{q}\,.
\end{eqnarray}

where we have used $\delta T_q/\bar{T}=\delta \rho_{\gamma q}/4\bar{\rho}_{\gamma}=\delta_{\gamma q}/3$. Nevertheless, for the ``delta'' contribution, $\dot{\tilde{\Psi}}_q$ and $\ddot{\tilde{\Psi}}_q$ satisfy the same relation than the standard case. Due to our decomposition, the tilde expresions are

\begin{eqnarray}
  \tilde{{\cal F}}(q)&=&-\frac{3}{2}\frac{A_q(t_{ls})}{q^2}(a_{DG}^2(t_{ls})\ddot{F}(t_{ls})+a_{DG}(t_{ls})\dot{a}_{DG}(t_{ls})\dot{F}(t_{ls}))\nonumber \\&+&\frac{\dot{\tilde{\Psi}}_q(t_{ls})}{q^2}\left(a_{DG}(t_{ls})\dot{a}_{DG}(t_{ls})-\frac{2}{3}\frac{a_{DG}^2(t_{ls})}{t_{ls}}\right)\,,\\
  \tilde{{\cal G}}(q)&=&-q\frac{\delta \tilde{u}_{\gamma q}(t_{ls})}{(1+3F(t_{ls}))a_{DG}(t_{ls})}+\frac{a_{DG}(t_{ls})\dot{\tilde{\Psi}}_q(t_{ls})}{q}\,.
\end{eqnarray}

Unfortunately, due to all the approximations we have used, we need to add some corrections to the solutions of the GR sector. After that, we will be able to find the numerical solutions for DG perturbations.\\
The first consideration is that in the set of equations presented in the matter era, we have used $R=3\bar{\rho}_B/4\rho_{\gamma}=0$, which is not valid in this era. Corrections to the solutions can be calculated using WKB approximation for perturbations\footnote{see Section 6.3: Scalar perturbations-long wavelengths, Cosmology, Weinberg.}\cite{weinberg2008cosmology}.
The second consideration that we must included in the solution of photons perturbations is the so-called Silk damping\footnote{see Section 6.4: Scalar perturbations-short wavelengths, Cosmology, Weinberg.}\cite{SILK1968,1983MNRAS.202.1169K}, which takes into account viscosity and heat conduction of the relativistic medium. Moreover, the transition from opaque to a transparent Universe at the last scattering moment was not instantaneous, but it could be considered a gaussian. This effect is known as Landau damping\footnote{see Section 7.2: Temperature multipole coefficients: Scalar modes, Cosmology, Weinberg.}. We must recall that the physical geometry now is described by $Y_{DG}(t)=a_{DG}(t)/a_{DG}(t=0)$, so the expression for both Silk and Landau effects have to be expressed in this geometry. With these considerations, the solutions of perturbations are given by:

\begin{eqnarray}\label{extended solutions}
  \dot{\Psi}_q(t_{ls})&=&-\frac{3q^2t_{ls}{\cal R}_q^o{\cal T}(\kappa)}{5a^2(t_{ls})}\,,\\
  \delta_{\gamma q}(t_{ls})&=&\frac{3{\cal R}_q^o}{5}\left[{\cal T}(\kappa)(1+3R_{ls})-(1+R_{ls})^{-1/4}e^{-q^2d^2_D/a_{ls}^2}\right.\nonumber\\&\times&\left.{\cal S}(\kappa)\cos\left(\int_0^{t_{ls}}\frac{qdt}{\sqrt{3(1+R(t))}a(t)}+\Delta(\kappa)\right)\right]\,,\\
  \delta u_{\gamma q}(t_{ls})&=&\frac{3{\cal R}_q^o}{5}\left[-t_{ls}{\cal T}(\kappa)+\frac{a(t_{ls})}{\sqrt{3}q(1+R_{ls})^{3/4}}e^{-q^2d^2_D/a_{ls}^2}\right.\nonumber\\&\times&\left.{\cal S}(\kappa)\sin\left(\int_0^{t_{ls}}\frac{qdt}{\sqrt{3(1+R(t))}a(t)}+\Delta(\kappa)\right)\right]\,,
\end{eqnarray}
Here we used an approximation given by $a_{DG}(t_{ls})\approx a(t_{ls})\propto t^{2/3}$, the error of this approximation is of the order $10^{-4}\%$. 

\begin{eqnarray}
  \dot{\Psi}_q(t_{ls})&=&-\frac{3q^2t_{ls}{\cal R}_q^o{\cal T}(\kappa)}{5a_{DG}^2(t_{ls})}\,,\\
  \delta_{\gamma q}(t_{ls})&=&\frac{3{\cal R}_q^o}{5}\left[{\cal T}(\kappa)(1+3R_{ls})-(1+R_{ls})^{-1/4}e^{-q^2d^2_D/a^2_{DG}(t_{ls})}\right.\nonumber\\&\times&\left.{\cal S}(\kappa)\cos\left(q\int_0^{t_{ls}}\frac{dt}{\sqrt{3(1+R(t))}a_{DG}(t)}+\Delta(\kappa)\right)\right]\,,\\
  \delta u_{\gamma q}(t_{ls})&=&\frac{3{\cal R}_q^o}{5}\left[-t_{ls}{\cal T}(\kappa)+\frac{a_{DG}(t_{ls})}{\sqrt{3}q(1+R_{ls})^{3/4}}e^{-q^2d^2_D/a_{DG}^2(t_{ls})}\right.\nonumber\\&\times&\left.{\cal S}(\kappa)\sin\left(q\int_0^{t_{ls}}\frac{dt}{\sqrt{3(1+R(t))}a_{DG}(t)}+\Delta(\kappa)\right)\right]\,,
\end{eqnarray}

where

\begin{eqnarray}\label{dampings}
  d^2_D=d^2_{Silk}+d^2_{Landau}\,,\\
  d^2_{Silk}=Y_{DG}^2(t_{ls})\int_0^{t_{ls}}\frac{t_{\gamma}}{6Y_{DG}^2(1+R)}\left\{\frac{16}{15}+\frac{R^2}{(1+R)}\right\}dt\,,\\
  d^2_{Landau}=\frac{\sigma_t^2}{6(1+R_{ls})}\,,
\end{eqnarray}

where $t_{\gamma}$ is the mean free time for photons and $R=3\bar{\rho}_B/4\bar{\rho}_{\gamma}=3h^2\Omega_BY_{DG}/4h^2\Omega_{\gamma}$.\\
In order to evaluate the Silk damping, we have
\begin{equation}
t_{\gamma}=\frac{1}{n_e\sigma_Tc}\,,
\end{equation}

where $n_e$ is the number density of electrons and $\sigma_T$ is the Thomson cross section. \\

On the other hand 
\begin{eqnarray}
    q\int_0^{r_{ls}} c_s dr&=& q\int_0^{t_{ls}}\frac{dt}{\sqrt{3(1+R(t))}a_{DG}(t)}\equiv qr^{SH}_{ls}\nonumber\\&=&\frac{q}{a_{DG}(t_{ls})}\cdot (a_{DG}(t_{ls})r_{ls}^{SH})=\frac{q}{a_{DG}(t_{ls})}\cdot d_H(t_{ls})
\end{eqnarray}

where $c_s$ is the speed of sound, $r^{SH}_{ls}$ is the sound horizon radial coordinate and $d_H$ is the horizon distance.\\

 With all this approximation, the transfers functions were simplified to the following expressions: 
 
\begin{eqnarray}
  {\cal F}(q)&=&\frac{1}{3}\delta_{\gamma q}(t_{ls})+\frac{a^2_{DG}(t_{ls})\dot{\Psi}_q(t_{ls})}{3q^2t_{ls}}\,,\\
  {\cal G}(q)&=&-q\frac{\delta u_{\gamma q}(t_{ls})}{(1+3F(t_{ls}))a_{DG}(t_{ls})}+\frac{a_{DG}(t_{ls})\dot{\Psi}_q(t_{ls})}{q}\,,
\end{eqnarray}

where $A_q(t_{ls})={\cal R}_q^o{\cal T}(\kappa)$. Then, wee replaced the GR solutions and we get

\begin{eqnarray}
{\cal F}(q)&=&\frac{{\cal R}_q^o}{5}\left[3{\cal T}(qd_T/a_{DG}(t_{ls}))R_{ls}-(1+R_{ls})^{-1/4}e^{-q^2d^2_D/a^2_{DG}(t_{ls})}\right.\nonumber\\&\times&\left.{\cal S}(qd_T/a_{DG}(t_{ls}))\cos\left(qd_H/a_{DG}(t_{ls})+\Delta(qd_T/a_{DG}(t_{ls}))\right)\right]\,,\\
 {\cal G}(q)&=&\frac{\sqrt{3}{\cal R}_q^o}{5(1+R_{ls})^{3/4}}e^{-q^2d^2_D/a^2_{DG}(t_{ls})}\nonumber\\&\times&{\cal S}(qd_T/a_{DG}(t_{ls}))\sin\left(qd_H/a_{DG}(t_{ls})+\Delta(qd_T/a_{DG}(t_{ls}))\right)\,,
\end{eqnarray}

where $\kappa=qd_T/a_{ls}$ (defined in eq. (\ref{kappa})) and

\begin{equation}
d_T(t_{ls})\equiv c\frac{\sqrt{2} a_{DG}(t_{ls})}{a_{EQ}H_{EQ}}  =c \frac{a_{DG}(t_{ls})\sqrt{\Omega_R}}{H_{0}\Omega_M}= c \frac{a_{DG}(t_{ls})}{100h}\sqrt{C(C+1)}\,.
\end{equation}

The final consideration that we must include is that due to the reionization of hydrogen at $z_{reion}=10$ by ultraviolet light coming from the first generation of massive stars, photons of the CMB have a probability of being scattered $1-\exp(-\tau_{reion})$. CMB has two contributions. The non-scattered photons provide the first contribution, where we have to correct by a factor given by $\exp(-\tau_{reion})$. The scattered photons provide the second contribution, but the reionization occurs at $z\ll z_L$ affecting only low $l$s. We are not interested in this effect, and therefore we will not include it. Measurements shows that in GR $\exp(-2\tau_{reion})\approx 0.8$.\\

On the other hand, we will use a standard parametrization of ${\cal R}_q^0$ given by

\begin{equation}
  |{\cal R}_q^0|^2=N^2q^{-3}\left(\frac{q/R_0}{\kappa_{\cal R}}\right)^{n_s-1}\,,
\end{equation}

where $n_s$ could vary with the wave number. It is usual to take $\kappa_{\cal R}=0.05$ {Mpc}$^{-1}$. 

Note that $d_A(t_{ls})=r_{ls}a_{DG}(t_{ls})$ is the angular diameter distance of the last scattering surface. 

\begin{eqnarray}
  d_A(t_{ls})&=&ca_{DG}(t_{ls})\int_{t_{ls}}^{t_0}\frac{dt'}{a_{DG}(t')}=c\frac{a_{DG}(t_0)}{1+z_{ls}}\int_{t_{ls}}^{t_0}\frac{dt'}{a_{DG}(t')}=c\frac{1}{1+z_{ls}}\int_{t_{ls}}^{t_0}\frac{dt'}{Y_{DG}(t')}\nonumber\\&=&c\frac{1}{1+z_{ls}}\int_{Y_{ls}}^{1}\frac{dY'}{Y_{DG}(Y')}\frac{dt}{dY'}=\frac{d_L(t_{ls})}{(1+z_{ls})^2}\,.
\end{eqnarray}
This is consistent with the luminosity distance definition\cite{Alfaro:2019qja}. Then, when we set $q=\beta l/r_{ls}$ we get 
\begin{eqnarray*}
  |{\cal R}_{\beta l/r_{ls}}^0|^2&=&N^2\left(\frac{\beta l}{r_{ls}}\right)^{-3}\left(\frac{\beta l}{\kappa_{\cal R}r_{ls}}\right)^{n_s-1} =N^2\left(\frac{\beta l}{r_{ls}}\right)^{-3}\left(\frac{\beta l a_{DG}(t_{ls})}{\kappa_{\cal R}r_{ls}a_{DG}(t_{ls})}\right)^{n_s-1}\\&=&N^2\left(\frac{\beta l}{r_{ls}}\right)^{-3}\left(\frac{\beta l a_{DG}(t_{ls})}{\kappa_{\cal R}d_A(t_{ls})}\right)^{n_s-1}\equiv N^2\left(\frac{\beta l}{r_{ls}}\right)^{-3}\left(\frac{\beta l}{l_R}\right)^{n_s-1}\,.
\end{eqnarray*}
Using a similar computations for the other distances, the final form of the form factors are given by
\begin{eqnarray}
{\cal F}(q)&=&\frac{{\cal R}_q^o}{5}\left[3{\cal T}(\beta l/l_T)R_{ls}-(1+R_{ls})^{-1/4}e^{-\beta^2l^2/l_D^2}\right.\nonumber\\&\times&\left.{\cal S}(\beta l/l_T)\cos\left(\beta l/l_H+\Delta(\beta l/l_T)\right)\right]\,,\\
 {\cal G}(q)&=&\frac{\sqrt{3}{\cal R}_q^o}{5(1+R_{ls})^{3/4}}e^{-\beta^2l^2/l_D^2}{\cal S}(\beta l/l_T)\sin\left(\beta l/l_H+\Delta(\beta l/l_T)\right)\,,
\end{eqnarray}
where
\begin{equation}
l_R=\frac{\kappa_{{\cal R}}d_A(t_{ls})}{a_{DG}(t_{ls})}\,,\;\; l_H=\frac{d_A(t_{ls})}{d_H(t_{ls})}\,,\;\; l_T=\frac{d_A(t_{ls})}{d_T(t_{ls})}\,,\;\; l_D=\frac{d_A(t_{ls})}{d_D(t_{ls})}\;.
\end{equation}

To summarize, for reasonably large values of $l$ (say $l>20$), CMB multipoles are given by 

\begin{eqnarray}\label{final multipole formula}
  \frac{l(l+1)C_{TT,l}^S}{2\pi}&=&\frac{4\pi T_0^2l^3\exp(-2\tau_{reion})}{r_{ls}^3}\int_1^{\infty}\frac{\beta d\beta}{\sqrt{\beta^2-1}}\nonumber\\
  &\times&\left[\left(F\left(\frac{l\beta}{r_{ls}}\right)+\tilde{F}\left(\frac{l\beta}{r_{ls}}\right)\right)^2+\frac{\beta^2-1}{\beta^2}\left(G\left(\frac{l\beta}{r_{ls}}\right)+\tilde{G}\left(\frac{l\beta}{r_{ls}}\right)\right)^2\right]\,.
\end{eqnarray}

Numerical solutions and other considerations should be included to compute the solution for the perturbations; however, this will be part of future work. It is remarkable the structure of eq. (\ref{final multipole formula}), where the delta sector contributes additively inside the integral. If we set all delta sector equal to zero, we recover the result directly for scalar temperature-temperature multipole coefficients in GR given by Weinberg. 

\section{Conclusions}
We discussed the implications of the first law of thermodynamics using the modified geometry of this model. We distinguished the physical densities from the GR densities in terms of which scale factor they dilute. However, knowing the solutions of the GR sector is enough for us to know about the behavior of the physical densities. Also, if we consider that the number of photons is conserved after the moment of decoupling, the black body distribution should keep the form, and that means that temperature is redshifted with the modified scale factor $Y_{DG}$. Finally, we stated the anzatz that the moment of equality between radiation and matter was the same in GR and in DG and we showed it implications in some parameters of the theory.\\
We had developed the theory of perturbations for Delta Gravity and its gauge transformations. Following Weinberg\cite{weinberg2008cosmology}, we used the Synchronous gauge which leaves a residual gauge transformation which can be used to set $\delta u_D=0$ (and also $\delta \tilde{u}_D=0$).\\
Then we computed the equations for cosmological perturbations using the hydrodynamic approximation, which we solved for the radiation era, while for a matter-dominated Universe, we presented the equations with the respective initial conditions. However, we did not solve them here because this will be part of a future work.\\
As in GR, we found an expression for temperature fluctuations in DG, studying the photon propagation in an effective metric, from the moment of the last scattering until now. We found that those temperature fluctuations can be split into three independent terms: an early term which only depends on the moment of the last scattering $t_{ls}$. An ISW term that includes the evolution of gravitational fields from the last scattering to the present and a late-term which depends on the actual value for those fields. We compute the gauge transformations which leaves ${\bf g}_{i0}=0$, and we found that those three terms are separately gauge invariants. Then, we derived the TT multipole coefficients for scalar modes, where we found that DG affects additively, which could have an observational effect that could be compared with Plank results and give a physical meaning for the so-called ``delta matter''.\\
With the full scalar expression for the CMB Power Spectrum coefficients, we can find the shape of the spectrum. In order to achieve it, we have to determine the best cosmological parameters that can describe the observational spectrum given by Planck \cite{BIB_PLANCK_Aghanim:2018eyx}. The determination of the cosmological parameters could be demanding (from a computational point of view), but if we constraints the cosmological parameters with the SNe-Ia analysis \cite{Universe5020051} the determination of the CMB Power Spectrum in DG could be more comfortable. In the context of the controversy about the $H_0$ value \cite{Riess2018} and other problems as the curvature measurements \cite{DiValentino2020} or the possibility of a Universe with less Dark Energy \cite{Kang_2020}, this work could provide an alternative to solve the today cosmological puzzle. Future work in this line is being carried out.

\acknowledgments
The author CR was supported by Conicyt PhD Fellowship No. 21150314, Fondecyt 1150390 and CONICYT-PIA-ACT14177. Marco San Martín was supported by Conicyt PhD Fellowship No. 21170604, Fondecyt 1150390 and CONICYT-PIA-ACT14177. J. Alfaro is partially supported by Fondecyt 1150390 and CONICYT-PIA-ACT14177.



\end{document}